\documentclass{emulateapj}

\shorttitle{Stellar Mass Function in MOIRCS Deep Survey}
\shortauthors{Kajisawa et al.}

\begin{document}


\title{MOIRCS Deep Survey. IV:\\ Evolution of Galaxy Stellar Mass Function back to $z\sim3$}


\author{M. Kajisawa\altaffilmark{1}, T. Ichikawa\altaffilmark{1}, I. Tanaka\altaffilmark{2}, M. Konishi\altaffilmark{3}, T. Yamada\altaffilmark{1}, M. Akiyama\altaffilmark{1},\\ R. Suzuki\altaffilmark{2}, C. Tokoku\altaffilmark{1}, Y. K. Uchimoto\altaffilmark{3}, T. Yoshikawa\altaffilmark{1}, M. Ouchi\altaffilmark{4}, I. Iwata\altaffilmark{5},\\ T. Hamana\altaffilmark{6}, M. Onodera\altaffilmark{7}}
%
\email{kajisawa@astr.tohoku.ac.jp}
%
%


\altaffiltext{1}{Astronomical Institute, Tohoku University, Aramaki,
Aoba, Sendai 980--8578, Japan}
\altaffiltext{2}{Subaru Telescope, National Astronomical Observatory
of Japan, 650 North Aohoku Place, Hilo, HI 96720, USA}
\altaffiltext{3}{Institute of Astronomy, University of Tokyo, Mitaka, Tokyo
181--0015, Japan}
\altaffiltext{4}{Observatories of the Carnegie Institution of Washington, 813 Santa Barbara Street, Pasadena, CA 91101, USA}
\altaffiltext{5}{Okayama Astrophysical Observatory, National Astronomical Observatory of Japan, Kamogata, Asakuchi, Okayama, 719--0232, Japan}
\altaffiltext{6}{National Astronomical Observatory of Japan, Mitaka, Tokyo
181--8588, Japan}
\altaffiltext{7}{Service d'Astrophysique, CEA Saclay, Orme des Merisiers, 91191 Gif-sur-Yvette Cedex, France}


\begin{abstract}
We use very deep near-infrared (NIR) imaging data  
obtained in MOIRCS Deep Survey (MODS) 
 to investigate the evolution of the galaxy stellar mass 
function back to $z\sim3$. 
The MODS data reach $J=24.2$, $H=23.1$, $K=23.1$ 
(5$\sigma$, Vega magnitude) over 103 arcmin$^2$ (wide) 
and $J=25.1$, $H=23.7$, 
$K=24.1$ over 28 arcmin$^2$ (deep) 
in the GOODS-North region.
The wide and very deep NIR data allow us to measure the number density of 
galaxies down to low stellar mass (10$^9$-10$^{10}$M$_{\odot}$) even 
at high redshift 
with high statistical accuracy.
The normalization of the mass function decreases  
with redshift and the integrated stellar mass density 
becomes $\sim$ 8--18\% of the local 
value at $z\sim2$ and $\sim$ 4--9\% at $z\sim3$, which are consistent with 
results of previous studies in general fields.
Furthermore, we found that the low-mass slope becomes steeper with redshift 
from $\alpha\sim -1.3$ at $z\sim 1$ to $\alpha\sim -1.6$ at $z\sim 3$, 
and that the evolution of the number density of low-mass 
(10$^9$-10$^{10}$M$_{\odot}$) galaxies is weaker than 
that of M$^*$ ($\sim$10$^{11}$M$_{\odot}$) galaxies.
This indicates that the contribution of low-mass galaxies to the total 
stellar mass density has been significant at high redshift.
The steepening of the low-mass slope with redshift is opposite 
trend expected from the stellar mass dependence of the specific 
star formation rate reported in previous studies. 
The present result suggests that 
the hierarchical merging process overwhelmed the 
effect of the stellar mass growth by star formation  
and was very important for the stellar mass 
assembly of these galaxies at $1\lesssim z \lesssim3$. 
\end{abstract}





\section{Introduction}

Understanding when and how galaxies built up their stellar mass is one of 
the most important issues in observational cosmology and galaxy formation.
Since the stellar mass of a galaxy is generally dominated by long-lived 
low-mass stars, it is considered as the integral of the
past star formation rate. Then, the evolution of the 
cosmic stellar mass density provides a global picture of the past history 
of star formation in the universe. Recent near-infrared (NIR) surveys 
allow us to measure the average stellar mass density of the universe
directly from low to high redshift (e.g., \citealp{wil08} and references therein). 
At $z\lesssim1$, many previous studies found the relatively mild evolution of 
the average stellar mass density, although there has been a significant 
evolution of the relative contributions from different morphological 
types or different SED/color types to the total stellar mass density 
(e.g., \citealp{bri00}, \citealp{dro04}, \citealp{bun05}, 
\citealp{pan06}, \citealp{bor06}, \citealp{fra06}, \citealp{ver08}, 
\citealp{ilb09}). 
In contrast, at $1\lesssim z \lesssim3$, a rapid evolution of the 
stellar mass density has been observed by studies based on deep surveys 
(\citealp{dic03}, \citealp{fon03}, \citealp{rud03}, 
\citealp{fon04}, \citealp{gla04}, \citealp{dro05}, \citealp{gwy05}, 
\citealp{rud06}, \citealp{fon06}, \citealp{arn07}, 
\citealp{poz07}, \citealp{per08}), and a significant fraction of the stellar mass 
in the present universe seems to have been formed at that epoch.
This is qualitatively consistent with results  
that the average star formation rate density in the 
universe peaked around $z\sim$ 1--3 (e.g., \citealp{mad96}, \citealp{hop04}, 
\citealp{hop06}), although several studies pointed out a 
quantitative discrepancy between the evolution of 
the cosmic star formation rate density and the stellar mass density 
(\citealp{rud03}, \citealp{rud06}, \citealp{hop06}, \citealp{per08}, 
\citealp{wil08}, see also \citealp{arn07}).

Stellar mass function (SMF) of galaxies and its evolution over cosmic time 
provide additional 
information on how stellar mass assembly of galaxies 
proceeded in different mass ranges, 
while the evolution of the integrated stellar mass density enables us to 
investigate when stars formed in the universe as mentioned above.
In paticular, the evolution of the number density of massive (e.g., 
M$_{\rm star}>10^{11}$M$_{\odot}$) galaxies has been extensively studied in 
order to address when massive galaxies assembled, which is considered as 
an important test for galaxy formation models (e.g., \citealp{gla04}, 
\citealp{cap06}, \citealp{con07}, \citealp{brt07}).   
While the comoving number density of these massive galaxies shows no 
significant evolution at $z\lesssim1$, it decreases significantly with 
redshift at $1\lesssim z \lesssim3$, although the observed number density 
of very massive galaxies 
at $z\sim$ 2--3 seems to be still higher than the predictions of the 
hierarchical galaxy formation models 
(\citealp{gla04}, \citealp{cap06}, \citealp{con07}).
In order to understand how the stellar mass assembly of massive galaxies 
proceeded, studies for the evolution of lower mass galaxies are also 
important, since many low-mass objects at high redshift 
are expected to merge over cosmic time 
to form massive galaxies in hierarchical structure formation scenarios.
The hierarchical merging process generally destroys small galaxies and builds
massive galaxies, and therefore this process is expected to 
 change the shape of the SMF, for example, to increase the characteristic 
mass of the Schechter function 
and/or to flatten the low-mass slope with time. 
In order to address the shape of the SMF reliably,  
it is essential to have 
a large sample of galaxies over a wide range of stellar mass. 
  
To investigate the evolution of galaxies over a wide range of mass is also 
important in the context of the `downsizing' scenario.
In the local universe, 
massive galaxies tend to have redder color and older stellar 
population, while most low-mass galaxies are young and actively star-forming 
(\citealp{kau03}, \citealp{bri04}). The star formation histories of galaxies 
 depend strongly on their stellar mass (e.g., \citealp{hea04}, \citealp{jim05}).
Therefore the histories of the stellar mass assembly of high and low-mass 
galaxies could also be different. For example, higher star formation rate 
relative to stellar mass 
(i.e., specific star formation rate) of low-mass galaxies is expected to 
lead to 
higher growth rate of their stellar mass than that of massive galaxies, 
if the effect of 
galaxy mergers is ignored. In fact, several studies reported that 
the most massive galaxies with 
M$_{\rm star}\sim10^{11.5}$-10$^{12}$M$_{\odot}$ show 
milder evolution in number density at $1\lesssim z \lesssim 3$ than 
galaxies with M$_{\rm star}\sim10^{11}$-10$^{11.5}$M$_{\odot}$, i.e., 
 the mass-dependent evolution of the number density of massive galaxies (\citealp{con07}, 
\citealp{ber07}, \citealp{per08}). At $z\lesssim$ 1--1.3, it was also 
observed that lower mass galaxies 
with M$_{\rm star}\sim10^{10}$-10$^{11}$M$_{\odot}$ evolve faster than  massive 
galaxies (\citealp{fon06}, \citealp{poz07}).

However, it is difficult to extend the 
investigation of galaxies over a wide range of stellar mass 
up to $z\sim$ 1--3, which is an important era in the histories of the stellar 
mass assembly of galaxies. 
This is because very deep NIR data are necessary to unbiasedly sample galaxies 
at high redshift down to low stellar mass, e.g., $10^{9}$-$10^{10}$M$_{\odot}$ 
 (\citealp{kaj06b}). 
NIR data sample the rest-frame optical to NIR for galaxies at $z\sim$ 1--3. 
Since NIR luminosity reflects total stellar mass of a 
galaxy relatively well (e.g., \citealp{bel01})  
and is not strongly affected by dust extinction, 
 NIR data are more suitable for construction of stellar mass limited samples  
than shorter wavelength data. NIR data are also important in that they straddle 
the redshifted  
Balmer/4000\AA\ break of these high-z galaxies, which is correlated strongly
 with the stellar 
M/L ratio and enables us to estimate stellar mass with high accuracy. 
Since such deep NIR data have been very limited so far,  
only a few studies investigated the evolution of the shape of the SMF up to 
$z\sim$ 1--3 using ultra deep NIR data with relatively small survey areas
(\citealp{gwy05}, \citealp{mar08}).

In this paper, we use very deep and wide 
NIR data obtained in MOIRCS Deep Survey (MODS; \citealp{kaj06a}, \citealp{ich07})
with publicly available multi-wavelength data in the Great
Observatories Origins Deep Survey (GOODS, \citealp{gia04}) in order to 
investigate the evolution of the SMF of galaxies back to 
$z\sim3$. 
The wide and deep NIR data allow us to construct a statistically large galaxy
sample over a wide range of stellar mass even at high redshift. 
Section 2 describes the data set and the procedures of source detection 
and photometry.
We present methods for constructing the SMF and investigate its evolution  
in section 3. In section 4, we compare the results with previous studies and 
discuss their implications for galaxy formation and evolution. 
A summary is provided in section 5.

We use a cosmology with H$_{\rm 0}$=70 km s$^{-1}$ Mpc$^{-1}$, 
$\Omega_{\rm m}=0.3$ and $\Omega_{\rm \Lambda}=0.7$.
The Vega-referred magnitude system is used throughout this paper.

\section{Observational data and source detection}
\label{sec:obs}


\begin{figure*}[t]
\begin{center}
\includegraphics[angle=0,scale=1.0]{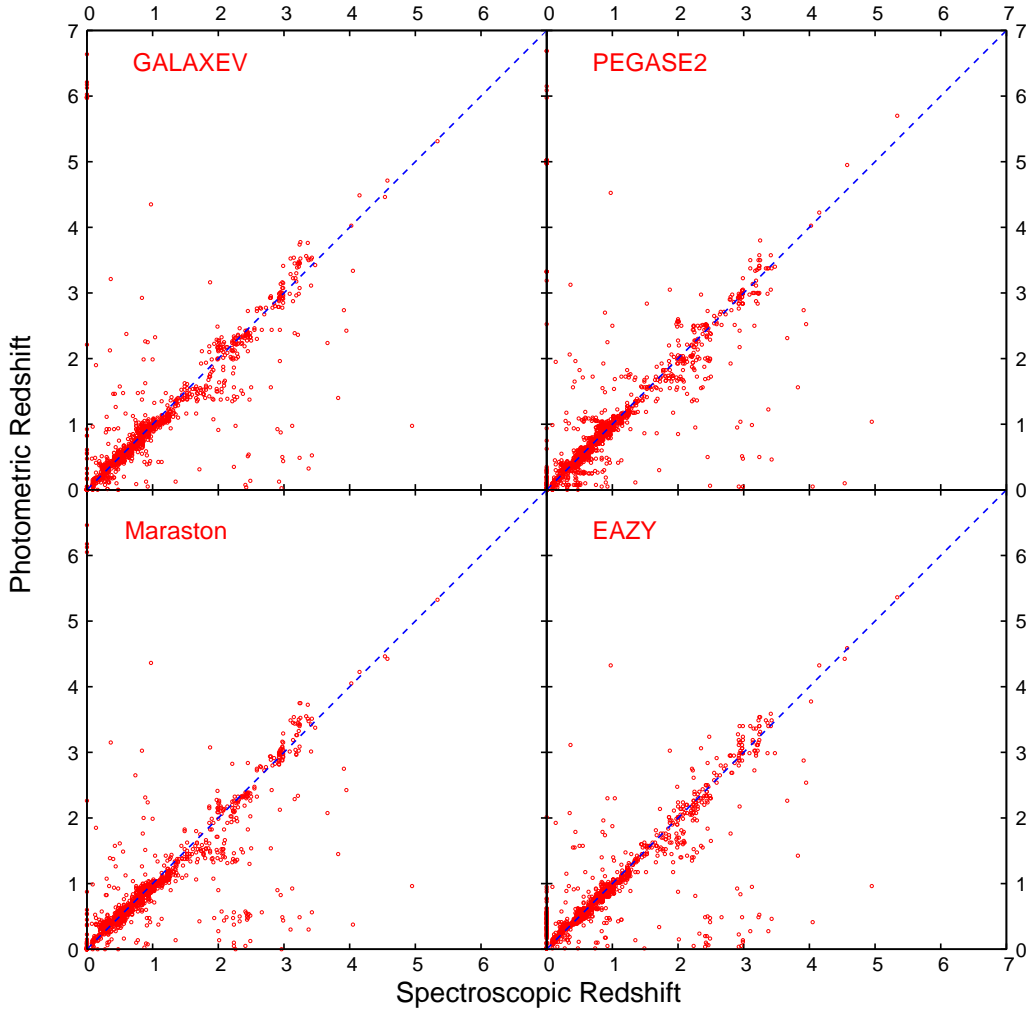}
\end{center}
\caption{Photometric redshift  vs.  spectroscopic redshift for galaxies 
with spectroscopic redshifts from the literature. Four panels show 
the different population synthesis models used as SED 
templates in the photometric redshift estimate. 
\label{fig:photz}}
\end{figure*}
\begin{figure*}[t]
\begin{center}
\includegraphics[angle=0,scale=1.0]{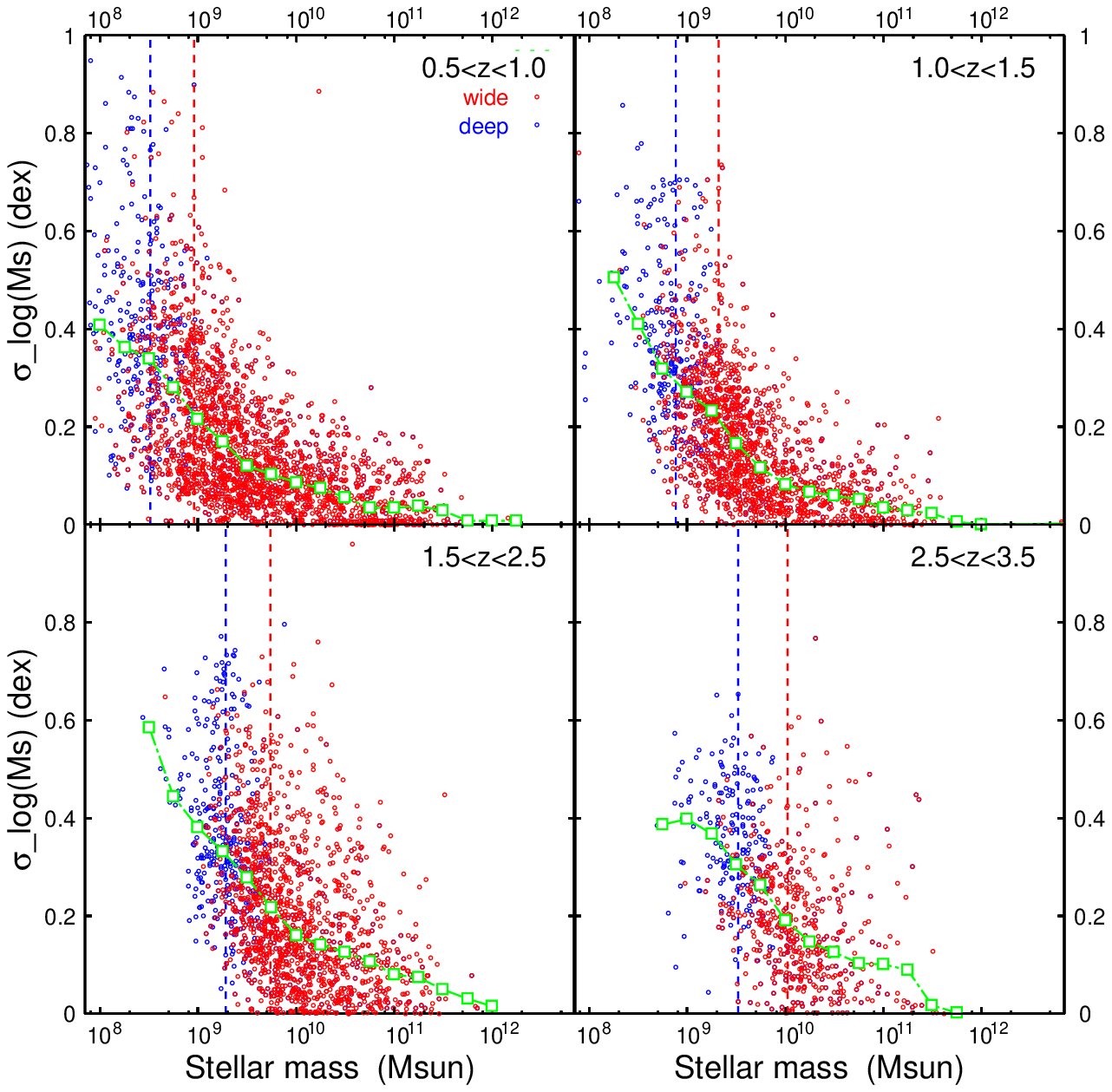}
\end{center}
\caption{Uncertainty of the estimated stellar mass as a function of stellar
mass in each redshift bin. 
Red circles represent the wide sample and blue circles show the deep sample.
Vertical dashed lines show the limiting stellar mass described in the text 
for the wide (red) and deep (blue) samples, respectively.
Open squares represent the median values at each stellar mass for 
all sample.
For objects without spectroscopic redshift,
the photometric redshift error is taken into account in the estimate of
stellar mass uncertainty (see text).
\label{fig:sigms}}
\end{figure*}

In the MOIRCS Deep Survey, we obtained the very deep $JHKs$-bands
 imaging data in the field of GOODS-North with 
Multi-Object InfraRed Camera and Spectrograph (MOIRCS, \citealp{ich06}, 
\citealp{suz08}) mounted on the Subaru Telescope.
The observations were carried out in the period 
from 2006 April to 2008 May.  
We also used archival MOIRCS data obtained by \cite{wan09} and \cite{bun09}.
Four MOIRCS pointings cover $\sim$ 70\% of the GOODS-North region ($\sim$ 103.3 
arcmin$^2$, 
 hereafter referred as `wide' field) 
and the total exposure time is 6.3--9.1 hours in $J$-band, 2.5-4.3 hours in 
$H$-band and 8.3-10.7 hours in $Ks$-band. 
One of the four pointings, which includes the Hubble Deep Field North (HDF-N, \citealp{wil96}), 
is also the ultra-deep field of the MODS (hereafter `deep' field),  
where the exposure time is 28.2 hours in $J$-band, 5.7 hours in 
$H$-band and 28.0 hours in $Ks$-band.

The data were reduced using a purpose-made IRAF-based software package 
called {\it MCSRED}
\footnote{http://www.naoj.org/staff/ichi/MCSRED/mcsred\_e.html}. 
The reduction procedures were as described in \cite{kaj06a}
except that a defringing process was additionally applied. 
A standard star P177-D \citep{leg06} was used for flux calibration.
In a photometric night, 
object frames for flux calibration were observed immediately after the 
standard star at similar airmass in all 4 pointings in $JHKs$-bands. 
Before combining the object frames, all frames were scaled to the count 
level of these calibration frames using relatively bright unsaturated sources.
We discarded the frames where the count level was very low (less than 70\% of 
the calibration frame) or the image quality was bad (FWHM of the PSF  
 larger than 1.2 arcsec for the wide field and 0.8 arcsec for the deep field).
The combined data reach  $J=24.2$, $H=23.1$ and $K=23.1$
(5$\sigma$, Vega magnitude) for the wide field and 
$J=25.1$, $H=23.7$ and 
$K=24.1$ for the deep field.  
Further details of the observations, 
reduction and quality of the data will be presented 
in a forthcoming paper (Kajisawa et al. in preparation).

The source detection was performed in the $Ks$-band image
using the SExtractor image analysis package (\citealp{ber96}).
We adopted MAG\_AUTO from the SExtractor as the total $K$-band magnitudes 
of the detected objects. In this study, we use
 the magnitude limited samples with $K<23$ and $K<24$ for the wide and 
deep fields, respectively. 
The detection completeness for point sources
 is more than 90\% at the $K$-band limits in the both 
samples and the false detection rate is expected to be less than $\sim$ 1\% 
 from the simulation of the source detection on 
the inverse $Ks$-band 
image with the same detection parameters (Kajisawa et al. in preparation).
We detected 6402 and 3203 sources above the $K$-band limits 
in the wide and deep fields, respectively.

In order to measure the optical-to-MIR spectral energy distributions 
(SEDs) of the 
detected objects, we used the publicly available multi-wavelength data 
in the GOODS 
field, namely KPNO/MOSAIC ($U$-band, \citealp{cap04}), HST/ACS 
($B$,$V$,$i$,$z$-bands, 
version 2.0 data, Giavalisco et al., in preparation, 
\citealp{gia04}) and Spitzer/IRAC (3.6$\mu$m, 4.5$\mu$m, 5.8$\mu$m, DR1 and DR2, 
Dickinson et al., in preparation), as well as the MOIRCS $J$ 
and $H$-bands images.
These multi-band images were aligned to the $Ks$-band image. 
The ACS and MOIRCS images were convolved to match the 
data with the poorest seeing for the both wide 
(to $\sim$ 0.6 arcsec FWHM) and deep 
(to $\sim$ 0.5 arcsec FWHM) samples. 
For the color measurements, we used a fixed aperture size with a 
diameter of 1.2 (1.0 for the deep sample) arcsec ($\sim2 \times$ seeing FWHM) 
for the ACS and MOIRCS data.
For MOSAIC ($U$-band) and IRAC (3.6$\mu$m, 4.5$\mu$m, 5.8$\mu$m) data, we first 
performed the aperture photometry of the wide(deep) sample 
with aperture sizes of 2.28(2.39), 2.89(3.07), 3.05(3.24), 3.80(4.02)  
arcsec, respectively, and then applied the aperture correction by using
 the light profiles of 
the $B$- and $Ks$-bands images smoothed to match 
the resolution of $U$-band and IRAC 
images, respectively (see Kajisawa et al. in preparation for details).

In the following analysis, we use objects which are detected above  
2$\sigma$ level in more than three bands ($Ks$-band and other two bands) 
because it is difficult to estimate the photometric redshift and stellar mass 
of those detected only in one or two bands.
In the wide and deep $K$-band magnitude limited samples, 
only 21/6402 and 42/3203 were excluded by this criterion, respectively.
As a result, 
the wide and deep samples consist of 6381 and 3161 objects, respectively, 
and the total number of objects used are 7563, in which 
1979 objects with $K<23$ 
in the deep field are included in the both samples.




\section{Analysis}
\subsection{Estimate of redshift and stellar mass}

In order to estimate the photometric redshift and stellar mass of the sample galaxies, 
we performed SED fitting of the multi-band 
photometry described above ($UBVizJHK$, 3.6$\mu$m, 4.5$\mu$m, 5.8$\mu$m) with 
population synthesis models. We adopted the standard minimum $\chi^2$ method for 
the fitting procedure.
The resulting best-fit redshift (i.e., photometric redshift) 
is used for objects without spectroscopic 
identifications and the best-fit stellar mass-to-luminosity (M/L) 
 ratio is used to calculate the stellar mass. 
We adopted spectroscopic redshifts (if available) 
from the literature (\citealp{coh00}, \citealp{coh01}, 
\citealp{daw01}, \citealp{wir04}, \citealp{cow04}, \citealp{tre05}, \citealp{cha05}, 
\citealp{red06}, \citealp{bar08}), including those from 
our NIR spectroscopic observation of $\sim$ 20 star-forming $BzK$ 
galaxies \citep{dad04} 
at $z\sim2$ with Subaru/MOIRCS (Yoshikawa et al. in preparation). 
For these objects,
the SED fitting was performed fixing the redshift to each spectroscopic value.

\begin{figure*}[t]
\begin{center}
\includegraphics[angle=0,scale=0.85]{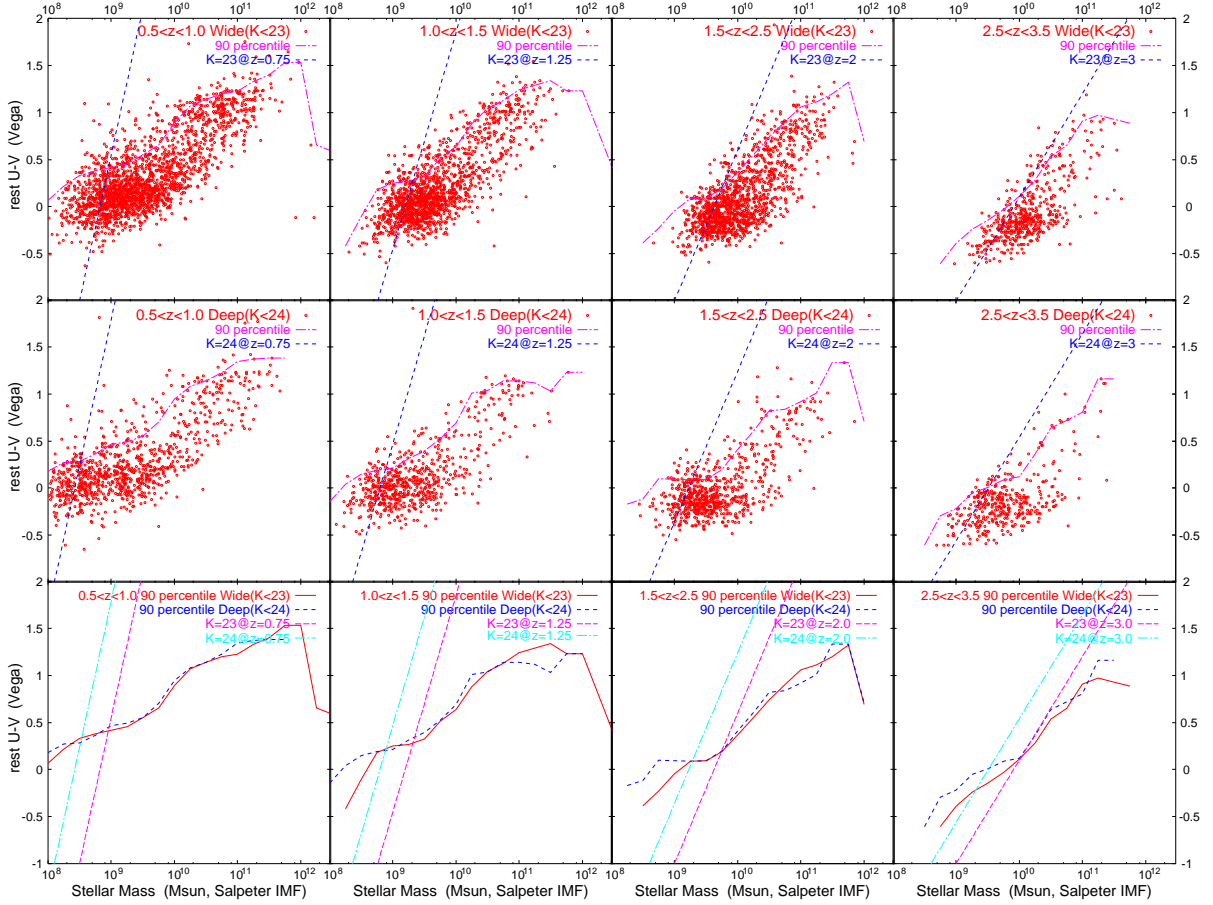}
\end{center}
\caption{{\bf Top and middle:} 
Rest-frame $U-V$ color distributions of wide (top) and deep (middle) 
samples for each redshift bin. Dashed lines represent the $K$-band magnitude limits 
($K=23$ for the wide sample and $K=24$ for the deep one) at the central redshift of 
each bin. Dashed-dot curves show 
the 90 percentile of $U-V$ color at each stellar mass.
{\bf Bottom:} Comparison of the 90 percentile of $U-V$ color between 
the wide (solid line) and deep (short dashed line) samples. 
Long-dashed line and dashed-dot line show the $K$-band magnitude limits for the 
wide and deep samples. 
\label{fig:uvms}}
\end{figure*}
\begin{figure*}[t]
\begin{center}
\includegraphics[angle=0,scale=1.1]{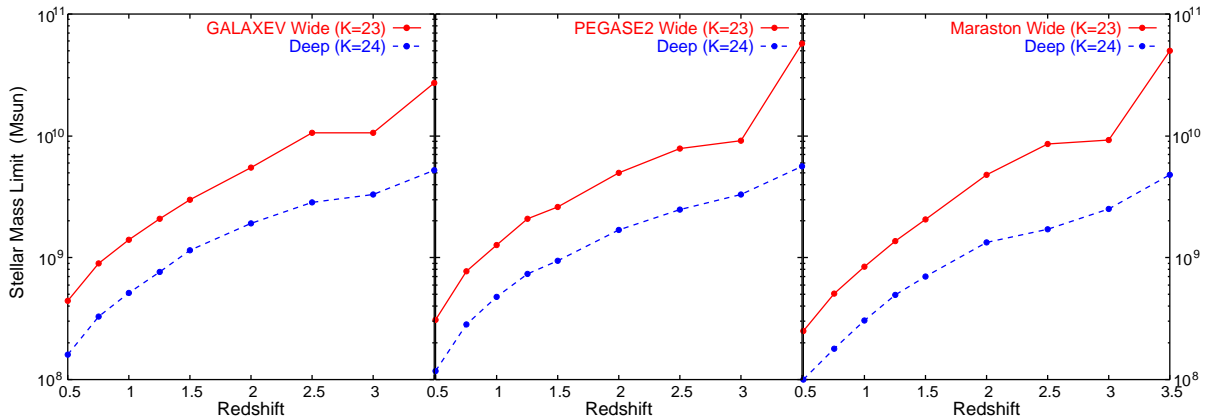}
\end{center}
\caption{Stellar mass limit as a function of 
redshift for the wide and deep samples. 
 Three panels show the results with the different SED models.
Solid lines show the wide sample and dashed lines show the deep one.
}
\label{fig:mlim}
\end{figure*}
\begin{figure*}[t]
\begin{center}
\includegraphics[angle=0,scale=0.85]{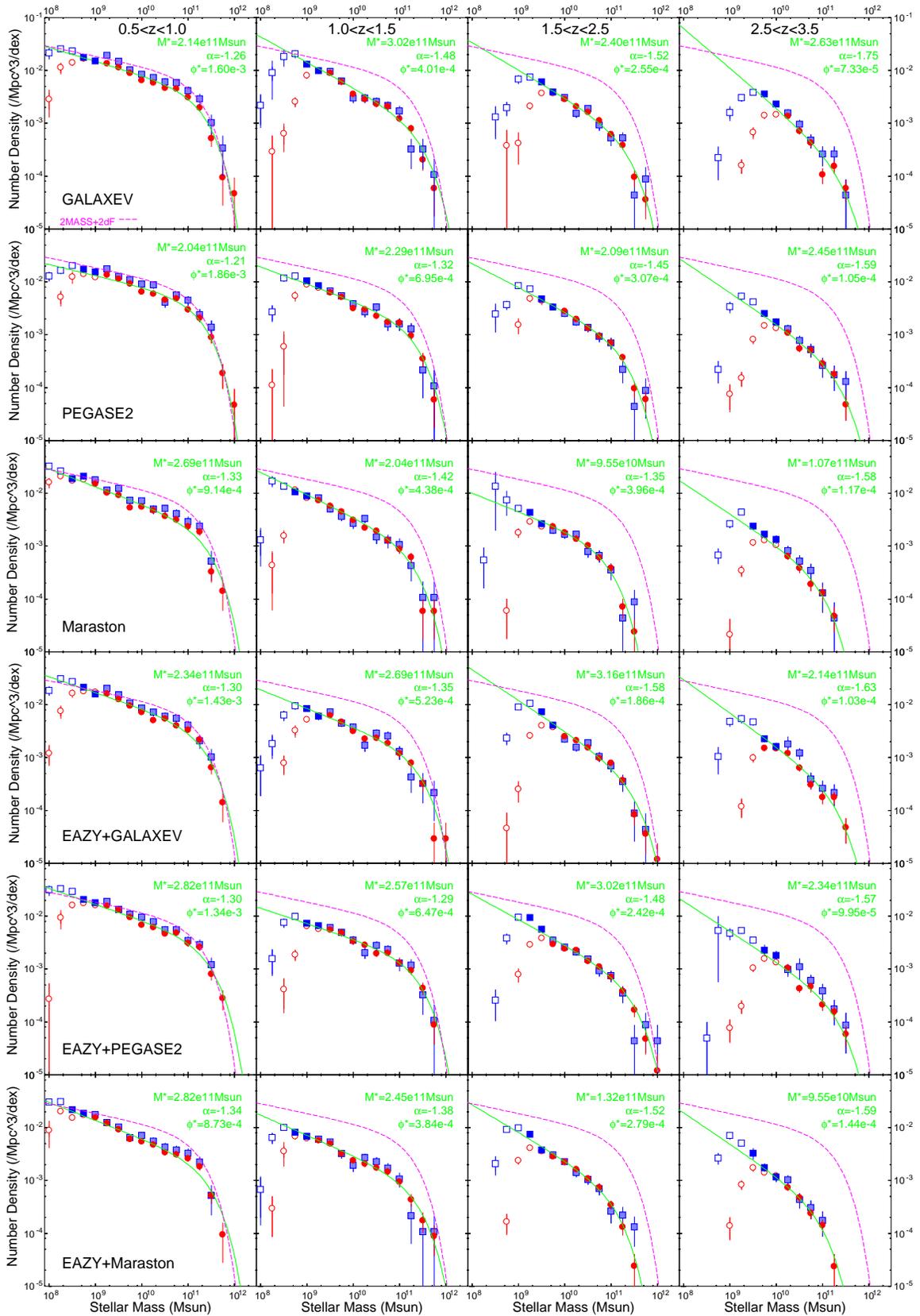}
\end{center}
\caption{Evolution of the stellar mass function of galaxies in the MODS field.
From top to bottom, 
the panels show the results with the different population 
synthesis models. Circles and squares show the SMF calculated with the 1/V$_{\rm max}$
 formalism for the wide and deep samples. Open symbols indicate data points 
located below the limiting stellar mass, where the incompleteness 
could be significant.
Solid symbols show data points above the limiting stellar mass.  
The results of the deep sample are plotted by shaded symbols
at stellar mass where the wide sample is also above the limiting mass.
Error bars are based on the Poisson statistics. 
The solid lines show the results 
calculated with the STY method for the all (wide and deep) samples. 
The best-fit Schechter parameters are also shown in each panel. 
For reference, the local SMF of \citet{col01} is shown as the dashed line.
}
\label{fig:mfeach}
\end{figure*}
\begin{figure*}[t]
\begin{center}
\includegraphics[angle=0,scale=1.0]{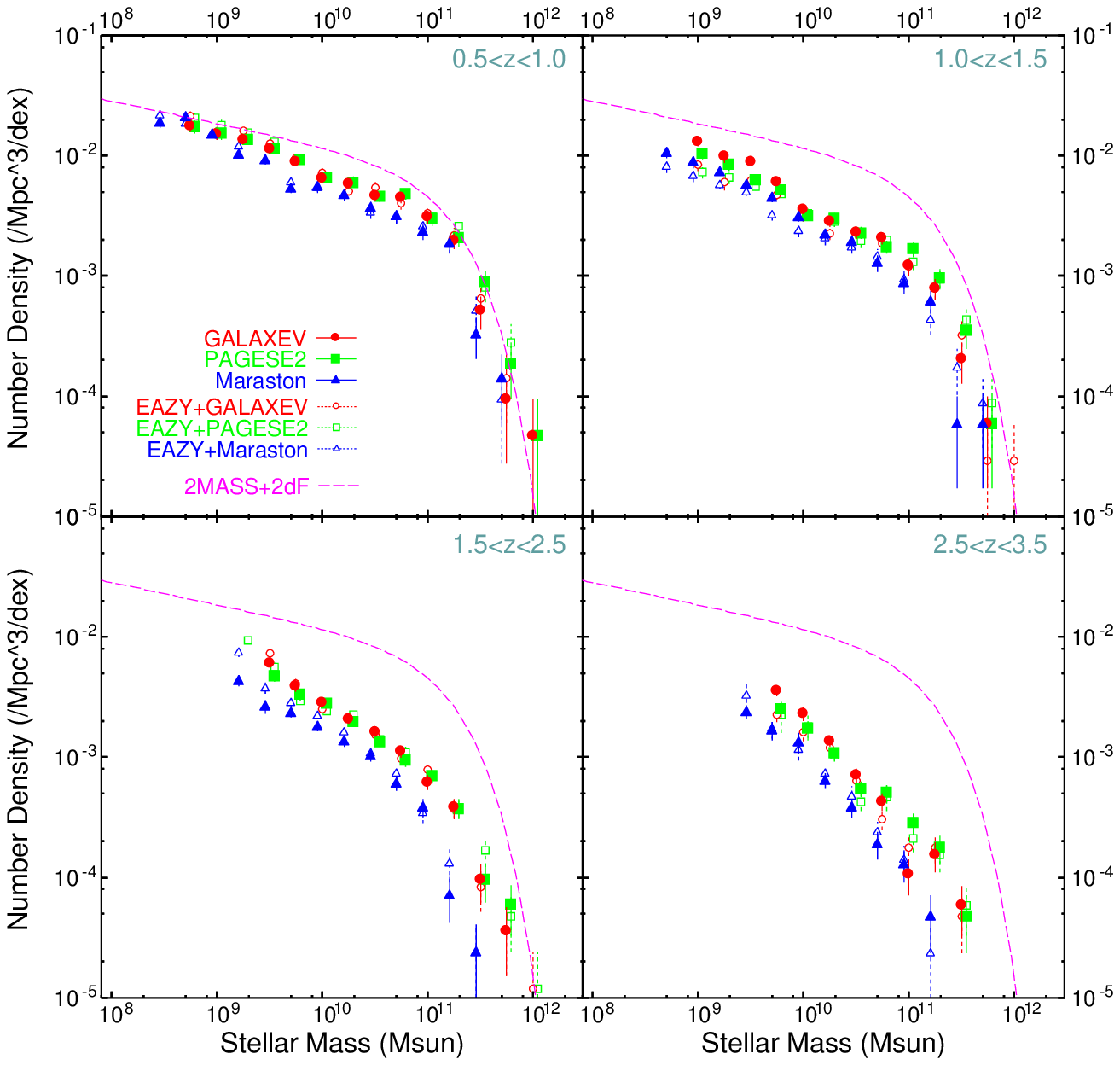}
\end{center}
\caption{Evolution of stellar mass function for different SED models. 
For each case, only data points above the limiting stellar mass are plotted. 
At stellar mass where the wide and deep samples 
are above the limiting mass, the data points 
for the wide sample are plotted (the same ones as 
solid symbols in Figure \ref{fig:mfeach}). 
Dashed line shows the local SMF of \citet{col01}.
}
\label{fig:mfall}
\end{figure*}

In this study, 
three models, i.e., GALAXEV (\citealp{bru03}),
 PEGASE version 2 (\citealp{fio97}) and \citet{mar05} model, 
were used as SED templates.
Comparisons among the results with different SED template sets allow us to 
check the systematic effects of these models on the estimate of redshift and stellar 
mass. 
We also 
used a public photometric redshift code, EAZY (\citealp{bra08}) for an independent check of the  
photometric redshift.
In all SED models, 
Salpeter IMF (\citealp{sal55}) with lower and upper mass limits of 0.1 and 100 M$_{\odot}$
is adopted for easy comparison among the results with the models and those in 
other studies. 
The details of the model templates from each population synthesis model
are the following. \\
(i) GALAXEV --- We used the similar template sets described in \citet{kaj05}.
Exponentially decaying star formation histories (SFHs) with the decaying 
timescale $\tau$ ranging between 0.1 and 20 Gyr are assumed. 
Calzetti extinction law
(\citealp{cal00}) in the range of $E(B-V)=$ 0.0--1.0 is adopted. 
Metallicity is changed from 
1/50 to one solar metallicity.\\
(ii) PEGASE2 --- We used the same template sets 
as those described in \citet{gra06}. 
Eight exponentially decaying SFHs with various gas infalling and star formation rate
timescales are assumed ($\tau_{\rm gas}=$ 0.1--5 Gyr and $\tau_{\rm SFR}=$ 0.1--20 Gyr, see Table 4 in \citealp{gra06} for details). The model includes 
the metallicity 
evolution and dust extinction in a self-consistent way. 
Following \citet{gra06}, 
we also added passively evolving templates with constant 
star formation rate (SFR) and several
truncation ages, 
 and dusty star-forming templates with a constant SFR and 
the Calzetti extinction with the range of $E(B-V)=$ 0.5--1.1.\\
(iii) Maraston (2005) --- The same SFHs and the same range of  
 the Calzetti extinction 
as in the case of GALAXEV are assumed ($\tau=$ 0.1--20 Gyr and $E(B-V)=$ 0.0--1.0).
Metallicity is changed from 0.5 to 2.0 solar metallicity. 
This model increases emphasis 
on contributions of thermally pulsating AGB stars to the SEDs.
In all these (i)-(iii) models, 
the model age is changed 
 from 50 Myr to the age of the universe at the observed redshifts.
We also added to the model SEDs the Lyman series absorption produced by the 
intergalactic medium following \citet{mad95}.\\
(iv) EAZY --- We used the default parameter setting and the default template set of 
{\it eazy\_v1.0}. The default template set consists of 5 principal component templates 
constructed from a large number of PEGASE models that cover a distribution of 
the SFH  of galaxies in the semi-analytic model by \citet{del07}.
In addition to these principal components, 
a dusty star-forming SED is added as a complementary template.
EAZY fits the multi-band photometry of the observed galaxies 
 with a linear combination of these templates (see \citealp{bra08} for details).
Since EAZY fits with the templates based on the principal components, we cannot 
derive the precise stellar M/L ratio from 
the fitting procedure.
In order to estimate the stellar mass, therefore, 
we performed SED fitting with the (i)-(iii) models,  
fixing the redshift to the output from EAZY. The results with the different population 
synthesis models at the same redshifts also enable us to check
the systematic effect of these models on the estimate of 
stellar M/L ratio independently.

In Figure \ref{fig:photz}, we compare the photometric redshifts with a sample of 
2102 spectroscopic redshifts from the literature described above.
The photometric redshift accuracy and the fraction of the
 catastrophic failure with $\delta$z/(1+z$_{\rm spec})>$0.5 for each SED model 
are ($\delta$z/(1+z$_{\rm spec})=-0.009\pm0.077$, 4.0\% outliers) for GALAXEV, 
($-0.002\pm0.103$, 4.0\%)  for PEGASE2, 
($-0.007\pm0.082$, 3.9\%)  for the Maraston model
and ($+0.003\pm0.098$, 4.7\%)  for EAZY, respectively.
 Although the photometric redshift accuracy is relatively good in all cases, there are some systematic
differences among
 the models especially at $1.5<z<3.0$ in Figure \ref{fig:photz}.
We check the effects of these differences of the photometric redshifts 
on the SMF in section \ref{sec:result}.
\begin{table}
\begin{center}
  \caption{Sample size in each redshift bin}
  \label{tab:num}
  \begin{tabular}{@{}lcccc   @{}}
  \tableline 
  \tableline
& redshift & model templates &  Wide\tablenotemark{a} & Deep\tablenotemark{a}\\ 
 \tableline
& $z=$0.5--1.0 & GALAXEV &  1945 (859) & 902 (311) \\
&             & PEGASE2 &  1860 (859) & 836 (311)  \\
&            & Maraston &  2118 (859) & 1023 (311) \\
&                & EAZY &  2141 (859) & 991 (311) \\
& $z=$1.0--1.5 & GALAXEV & 1426 (353) & 635 (105) \\
&             & PEGASE2 & 1319 (353) & 608 (105) \\
&            & Maraston & 1403 (353) & 573 (105) \\
&                & EAZY & 1141 (353) & 464 (105) \\
& $z=$1.5--2.5 & GALAXEV & 1306 (209) & 666 (75) \\
&             & PEGASE2 & 1380 (209) & 677 (75) \\
&            & Maraston & 1015 (209) & 488 (75) \\
&                & EAZY & 1334 (209) & 774 (75) \\
& $z=$2.5--3.5 & GALAXEV & 487 (95) & 366 (57) \\
&             & PEGASE2 & 507 (95) & 370 (57) \\
&            & Maraston & 427 (95) & 302 (57) \\
&                & EAZY & 494 (95) & 356 (57) \\
\tableline
\label{table:sample}
\end{tabular}
\vspace{-6mm}
\tablenotetext{a}{Number in the parenthesis indicates objects with spectroscopic redshift.}
\end{center}
\end{table}

Table \ref{table:sample} lists the number of objects in each redshift bin for 
different SED models. 
These redshift bins, of which width is sufficiently larger than 
the typical photometric redshift errors, 
were defined so as to include a reasonable 
number of galaxies in the bin for 
calculating the SMF.  
The comoving volumes are $8.5\times10^{4}$ Mpc$^{3}$ 
($2.5\times10^{4}$ Mpc$^{3}$ for the deep field) at $0.5<z<1.0$, 
$1.4\times10^{5}$ Mpc$^{3}$ ($3.7\times10^{4}$ Mpc$^{3}$) at $1.0<z<1.5$, 
$3.4\times10^{5}$ Mpc$^{3}$ ($9.2\times10^{4}$ Mpc$^{3}$) at $1.5<z<2.5$ and 
$3.4\times10^{5}$ Mpc$^{3}$ ($9.3\times10^{4}$ Mpc$^{3}$) at $2.5<z<3.5$, 
respectively.
The parenthesis in Table 
\ref{table:sample} represents the number of objects with spectroscopic redshift.
The fraction of spectroscopically identified sources is relatively high, 
thanks to the extensive spectroscopic surveys in this field.






As discussed in previous studies (e.g., \citealp{pap01}, 
\citealp{kaj05}, \citealp{sha05}), the uncertainty of 
the stellar mass is smaller than other parameters such as 
stellar age, star formation timescale, metallicity and dust extinction.
While the broad-band SEDs are degenerated with respect to these parameters of 
the stellar populations (stellar age, star formation timescale, 
metallicity and dust extinction), the stellar M/L ratio 
(and therefore 
stellar mass) is much less affected by the degeneracy because the effects of 
these parameters on the stellar M/L ratio tend to be canceled out with each 
other.
Figure \ref{fig:sigms} shows the uncertainty of the stellar mass estimated from 
a $\chi^2$ map in SED fitting with the GALAXEV model for each object.
For sources without spectroscopic identifications, we also varied redshift as a
 free parameter in the calculation of the $\chi^2$ map to take into account 
the photometric redshift error.
The uncertainty increases with decreasing stellar mass and increasing 
redshift. The stellar mass errors become $\sim$ 0.3--0.4 dex at the limiting 
stellar mass 
(the vertical dashed lines in the figure) described in the next 
subsection.

\subsection{Stellar mass-limited sample}
\label{sec:sample}

The $K$-band magnitude limited sample does not have a sharp limit in stellar 
mass even at a fixed redshift, because the stellar M/L ratio 
at the observed $K$-band varies with different stellar populations. 
We used the rest-frame $U-V$ color distribution as a function of 
stellar mass in each redshift bin 
to estimate the limiting stellar mass above which most of galaxies are expected 
to be brighter than the magnitude limits and detected in the $Ks$-band image.
Since rest-frame $U-V$ is correlated with stellar M/L ratio 
(e.g., \citealp{rud03}, \citealp{mar07}), one can predict 
the mass dependence of stellar M/L and then estimate 
 the effect of magnitude limit on the stellar mass distribution.

Figure \ref{fig:uvms} shows the rest-frame $U-V$ distribution 
in each redshift bin 
for the wide sample (top panels) and the deep (middle panels) sample.
A dashed line in each panel represents 
the $K$-band magnitude limit ($K=23$ for the wide sample
 and $K=24$ for the deep one). 
All objects with stellar mass larger than this line 
(on the right side of the line in the figure) at each $U-V$ value are 
brighter than the 
magnitude limit. In order to calculate the line, we used the stellar 
M/L ratio and the
rest-frame color of the GALAXEV model with various 
star formation histories, dust extinction, metallicity. 
The maximum mass 
was selected from the possible range of the stellar mass for the models 
with $K=23$ (or 24 for the deep sample) and each $U-V$ color 
in order to depict the dashed line in the figure.
Dashed-dot lines in Figure \ref{fig:uvms} represent the 90 percentile
 of $U-V$ color at each stellar mass. 
We adopted the point where the lines of the magnitude limit and  
90 percentile of $U-V$ color cross with each other as the limiting stellar mass.
Above this limiting mass, more than 90\% of objects are expected to be 
brighter than the $K$-band magnitude limit.

The bottom panels in Figure \ref{fig:uvms} show the comparison of the 90 percentiles of $U-V$ 
color for the wide and deep samples. 
The 90 percentiles for the both samples agree well with each other even near  
the limiting mass of the wide sample. 
This suggests that the incompleteness near the magnitude limit of the wide 
sample does not strongly affect the $U-V$ color distribution, although some red 
galaxies might be missed on the left side of the long-dashed line. 

Figure \ref{fig:mlim} shows the calculated limiting stellar mass as a function of redshift
for the wide and deep samples. We also estimated these mass limits 
for the cases with 
PEGASE2 and Maraston models as well as the case with GALAXEV.
As seen in Figure \ref{fig:uvms} and other previous studies (e.g., \citealp{kaj05}, 
\citealp{kaj06b}, \citealp{lab05}, \citealp{tay09}), 
less massive galaxies tend to have bluer rest-frame 
color even at high redshift. 
Such a mass-dependent color distribution can be seen 
well above the $K$-band magnitude limit up to at least $z\sim 2.5$ in Figure 
\ref{fig:uvms}.
Since the bluer color of low-mass galaxies indicates a lower M/L ratio, 
we can detect galaxies down to the relatively lower mass limit 
with high completeness compared with, for example, 
the mass limit based on the M/L ratio of 
the passively evolving models, which is used in other previous studies 
(e.g., \citealp{dic03}, \citealp{fon04}).
On the other hand, our 'wide' field data 
are relatively shallow for galaxies at $z>3$ 
(top right panel in Figure \ref{fig:uvms}) and the completeness is
 relatively low 
even at high mass, where galaxies tend to have red rest-frame colors 
(high M/L ratios).
We use objects with the stellar mass larger than these mass limits at each 
redshift to estimate and discuss the SMF in the following.
\begin{figure*}[t]
\begin{center}
\includegraphics[angle=0,scale=0.8]{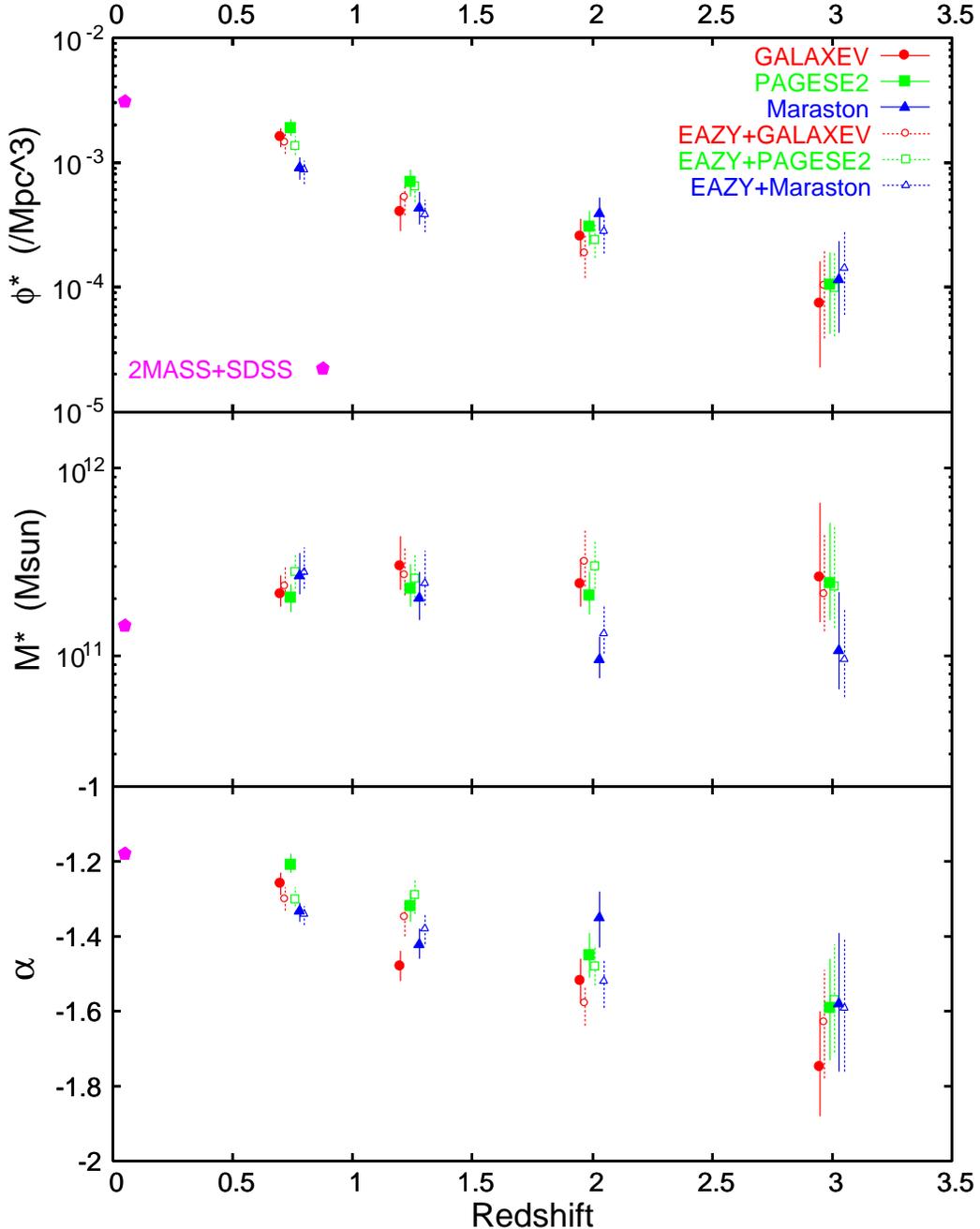}
\end{center}
\caption{Best-fit Schechter parameters as a function of redshift.
Different symbols represent the results with
 different population synthesis models. 
Data points for different SED models are plotted with small horizontal 
offsets for clarity. Those of the local SMF of \citet{col01} are also 
shown for reference.
}
\label{fig:schpara}
\end{figure*}
\begin{table*}[t]
\begin{center} 

\caption{Best-fit Schechter parameters obtained with the STY method}
  \begin{tabular}{@{}llcccc  @{}}
  \tableline 
  \tableline
&SED model& redshift bin & $\alpha$ & $\log_{10}$M$^*$(M$_{\odot}$) & $\log_{10}\phi^*$ (Mpc$^{-3}$)\\ 
\tableline 
& GALAXEV & $z=$ 0.5--1.0 &-1.26 $_{-0.03}^{+0.03}$ &11.33 $_{-0.07}^{+0.10}$ & -2.79 $_{-0.08}^{+0.07}$ \\
&& $z=$ 1.0--1.5 &-1.48 $_{-0.04}^{+0.04}$ &11.48 $_{-0.13}^{+0.16}$ & -3.40 $_{-0.15}^{+0.13}$ \\
&& $z=$ 1.5--2.5 &-1.52 $_{-0.06}^{+0.06}$ &11.38 $_{-0.12}^{+0.14}$ & -3.59 $_{-0.16}^{+0.14}$ \\
&& $z=$ 2.5--3.5 &-1.75 $_{-0.13}^{+0.15}$ &11.42 $_{-0.24}^{+0.40}$ & -4.14 $_{-0.51}^{+0.34}$ \\
\tableline
& PEGASE2 & $z=$ 0.5--1.0 &-1.21 $_{-0.02}^{+0.03}$ &11.31 $_{-0.08}^{+0.07}$ & -2.73 $_{-0.06}^{+0.07}$ \\
&& $z=$ 1.0--1.5 &-1.32 $_{-0.04}^{+0.04}$ &11.36 $_{-0.10}^{+0.13}$ & -3.16 $_{-0.11}^{+0.10}$ \\
&& $z=$ 1.5--2.5 &-1.45 $_{-0.06}^{+0.06}$ &11.32 $_{-0.10}^{+0.13}$ & -3.51 $_{-0.15}^{+0.12}$ \\
&& $z=$ 2.5--3.5 &-1.59 $_{-0.14}^{+0.13}$ &11.39 $_{-0.20}^{+0.32}$ & -3.98 $_{-0.40}^{+0.26}$ \\
\tableline
& Maraston & $z=$ 0.5--1.0 &-1.33 $_{-0.03}^{+0.02}$ &11.43 $_{-0.10}^{+0.12}$ & -3.04 $_{-0.10}^{+0.08}$ \\
&& $z=$ 1.0--1.5 &-1.42 $_{-0.04}^{+0.04}$ &11.31 $_{-0.12}^{+0.14}$ & -3.36 $_{-0.13}^{+0.12}$ \\
&& $z=$ 1.5--2.5 &-1.35 $_{-0.08}^{+0.07}$ &10.98 $_{-0.10}^{+0.12}$ & -3.40 $_{-0.14}^{+0.12}$ \\
&& $z=$ 2.5--3.5 &-1.58 $_{-0.18}^{+0.19}$ &11.03 $_{-0.21}^{+0.31}$ & -3.93 $_{-0.43}^{+0.30}$ \\
\tableline
& EAZY $+$ GALAXEV & $z=$ 0.5--1.0 &-1.30 $_{-0.03}^{+0.03}$ &11.37 $_{-0.08}^{+0.10}$ & -2.84 $_{-0.08}^{+0.07}$ \\
&& $z=$ 1.0--1.5 &-1.35 $_{-0.05}^{+0.04}$ &11.43 $_{-0.11}^{+0.15}$ & -3.28 $_{-0.14}^{+0.11}$ \\
&& $z=$ 1.5--2.5 &-1.58 $_{-0.06}^{+0.05}$ &11.50 $_{-0.13}^{+0.18}$ & -3.73 $_{-0.20}^{+0.15}$ \\
&& $z=$ 2.5--3.5 &-1.63 $_{-0.15}^{+0.14}$ &11.33 $_{-0.20}^{+0.32}$ & -3.99 $_{-0.42}^{+0.28}$ \\
\tableline
& EAZY $+$ PEGASE2 & $z=$ 0.5--1.0 &-1.30 $_{-0.02}^{+0.03}$ &11.45 $_{-0.09}^{+0.09}$ & -2.87 $_{-0.07}^{+0.08}$ \\
&& $z=$ 1.0--1.5 &-1.29 $_{-0.05}^{+0.04}$ &11.41 $_{-0.10}^{+0.13}$ & -3.19 $_{-0.13}^{+0.09}$ \\
&& $z=$ 1.5--2.5 &-1.48 $_{-0.05}^{+0.04}$ &11.48 $_{-0.12}^{+0.14}$ & -3.62 $_{-0.15}^{+0.14}$ \\
&& $z=$ 2.5--3.5 &-1.57 $_{-0.14}^{+0.15}$ &11.37 $_{-0.22}^{+0.32}$ & -4.00 $_{-0.39}^{+0.28}$ \\
\tableline
& EAZY $+$ Maraston & $z=$ 0.5--1.0 &-1.34 $_{-0.03}^{+0.02}$ &11.45 $_{-0.09}^{+0.13}$ & -3.06 $_{-0.11}^{+0.07}$ \\
&& $z=$ 1.0--1.5 &-1.38 $_{-0.04}^{+0.04}$ &11.39 $_{-0.12}^{+0.17}$ & -3.42 $_{-0.14}^{+0.12}$ \\
&& $z=$ 1.5--2.5 &-1.52 $_{-0.07}^{+0.06}$ &11.12 $_{-0.11}^{+0.14}$ & -3.55 $_{-0.17}^{+0.13}$ \\
&& $z=$ 2.5--3.5 &-1.59 $_{-0.17}^{+0.18}$ &10.98 $_{-0.20}^{+0.27}$ & -3.84 $_{-0.38}^{+0.28}$ \\
\tableline
\label{tab:schpara}
\end{tabular}
\end{center}
\end{table*}

\subsection{Deriving the stellar mass function}
The SMF of galaxies was derived with the non-parametric
1/V$_{\rm max}$ formalism and the parametric STY method \citep{san79}.
Both methods are commonly used to estimate the luminosity function and stellar mass
function of galaxies.

In the V$_{\rm max}$ method, 
V$_{\rm max}$ was calculated with the best-fit model SED template for each galaxy.
For each best-fit SED, we estimated the $K$-band apparent magnitude as a function of 
redshift, taking into account of both the luminosity distance and $K$-correction. 
Then we determined the maximum redshift, 
$z_{\rm max}$ above which the object becomes fainter than the $K$-band 
magnitude limit ($K=23$ for the wide sample or $K=24$ for the deep sample).
V$_{\rm max}$ is a comoving volume 
integrated from the lower limit of each redshift bin to $z_{\rm max}$ or the upper 
limit of the bin (the smaller of these two).
Then 1/V$_{\rm max}$ estimates were used to calculate 
the number density of galaxies in each mass bin.

In the STY method, assuming the Schechter function form \citep{sch76} 
for the SMF, 
we estimated best-fit values of the Schechter parameters ($\alpha$, 
M$^*$, $\phi^{*}$). 
The limiting stellar mass M$_{\rm lim}$(z) 
(described in the previous subsection) 
for the redshift of each object 
was used to calculate the probability that 
the object has the observed stellar mass M as 
\begin{equation}
p=\frac{\phi(M)}{\int^{\infty}_{M_{\rm lim}(z)}\phi(M) dM}
\end{equation}
Here $\phi$(M) is the SMF represented by the Schechter function.
We searched for the values of the Schechter parameters ($\alpha$, 
M$^*$, $\phi^{*}$) which maximize the likelihood L$=\prod p$, the products of 
the probability densities for the objects with the stellar mass larger than 
M$_{\rm lim}$(z) in each redshift bin.
Both the wide and deep samples were used simultaneously in the maximum 
likelihood technique. 
\begin{figure*}[t]
\begin{center}
\includegraphics[angle=0,scale=1.0]{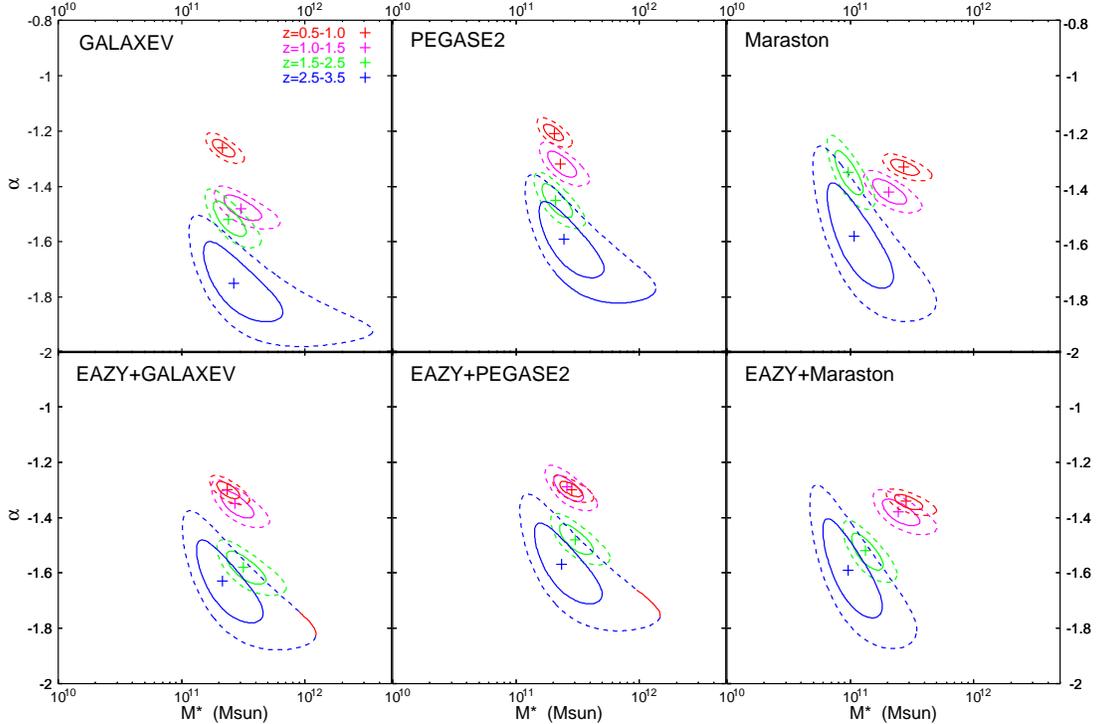}
\end{center}
\caption{Evolution of the Schechter parameters in 
M$^*$--$\alpha$ plane for the different SED models.
Crosses show the best-fit values determined with the STY method.
 1$\sigma$ (solid) and 2$\sigma$ (dashed) error contours are also shown. 
}
\label{fig:malpha}
\end{figure*}

\subsection{Evolution of the stellar mass function}
\label{sec:result}

Figure \ref{fig:mfeach} shows the stellar mass function (SMF) 
of galaxies in the 
different redshift bins for the different SED models. 
The results of the V$_{\rm max}$ method and 
the best-fit Schechter function estimated 
with the STY method are plotted in each panel.
Error bars are based on the Poisson statistics. 
Dashed lines show the local SMF derived from the 2dF and 2MASS surveys 
\citep{col01} 
with the small correction for the ``maximum age'' method as described in 
\citet{fon04}.
In Figure \ref{fig:mfall}, 
we plot the combined wide and deep complete data 
(same as the solid symbols in Figure \ref{fig:mfeach}) for the 
different SED models with different symbols in the same panel. 

Figures \ref{fig:mfeach} and \ref{fig:mfall} show that 
the SMFs obtained from 
 different samples and SED models are in good agreement,  
although there are some systematic differences among the SMFs. 
The number densities for the deep sample are systematically larger 
(by $\sim$ 0.1 dex) 
than those for the wide sample at $0.5<z<1.0$ for all SED models.
Our deep field is centered at the HDF-N field where the extensive spectroscopic 
surveys revealed the large scale structures at $z=0.85$ and $z=1.02$ 
(\citealp{coh00}, \citealp{wir04}). 
These filaments or clumps 
around the HDF-N could cause the slightly larger number density in our deep
field, and such differences can be considered as the possible 
field-to-field variance. Slighter excess in the deep field 
can be also seen at $2.5<z<3.5$ in most SED-model cases, 
and this may also be due to large scale structures.
On the other hand, the systematic differences among the different SED models
in Figure \ref{fig:mfall} seem to be larger. The number density of galaxies 
for the Maraston model is systematically smaller 
by $\sim$ 0.15--0.2 dex (up to $\sim$ 0.5 dex at $>10^{11}$M$_{\odot}$)
than those for 
the GALAXEV and PEGASE2 models especially at $z>1.5$. 
Since the differences are also seen between the EAZY$+$Maraston and 
EAZY$+$GALAXEV/PEGASE2, where the same redshifts of EAZY are used, 
the differences of the estimated stellar M/L ratio among the SED models 
cause the systematic differences of the number density.
For example, 
\citet{mar06} performed the broad-band SED fitting of relatively young 
($\sim$ 0.2--2 Gyr stellar age) galaxies at $1.4<z<2.7$ with the Maraston and 
GALAXEV models and reported that the Maraston model gives systematically 
younger age and lower stellar mass 
($\sim$ 60\%) than the GALAXEV model. Such difference of 
the estimated stellar mass seems to explain the differences seen in Figure 
\ref{fig:mfall}. If the fraction of young galaxies becomes larger 
at high redshift, the larger differences in the SMFs at $z>1.5$ could be also 
explained because the contribution of TP-AGB stars is expected to be 
significant 
in the relatively young ages ($\sim$ 0.2--2 Gyr, \citealp{mar05}, \citealp{mar06}).

\begin{figure*}[t]
\begin{center}
\includegraphics[angle=0,scale=1.5]{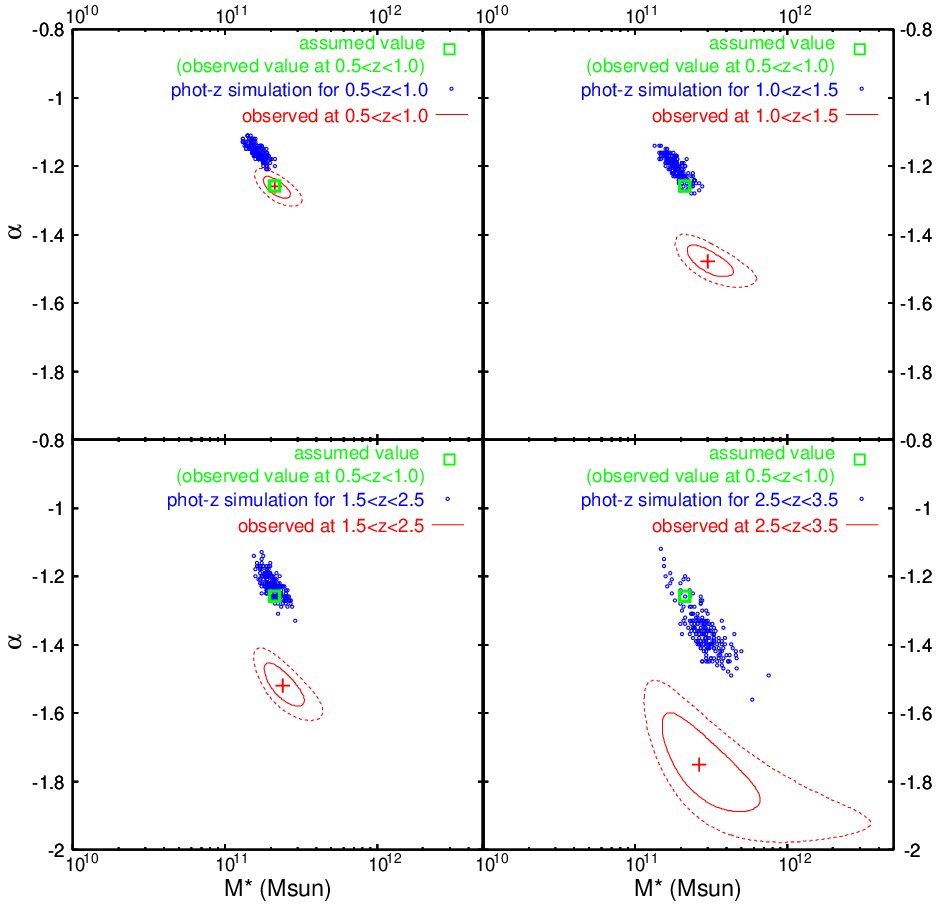}
\end{center}
\caption{Monte Carlo 
simulation for the effects of photometric redshift uncertainty on
the shape (M$^*$ and $\alpha$) of stellar mass function with 
the GALAXEV model.
Large square in each redshift bin shows 
the observed values at $0.5<z<1.0$ and these are 
assumed not to evolve with redshift in the simulation. 
Small circles show the results of 200 simulations (see text for details).
Observed M$^*$--$\alpha$ values (cross) and 1$\sigma$ (solid) and 2$\sigma$ 
(dashed) error contours are also shown for each redshift bin.
}
\label{fig:masim}
\end{figure*}
\begin{figure*}[t]
\begin{center}
\includegraphics[angle=0,scale=1.0]{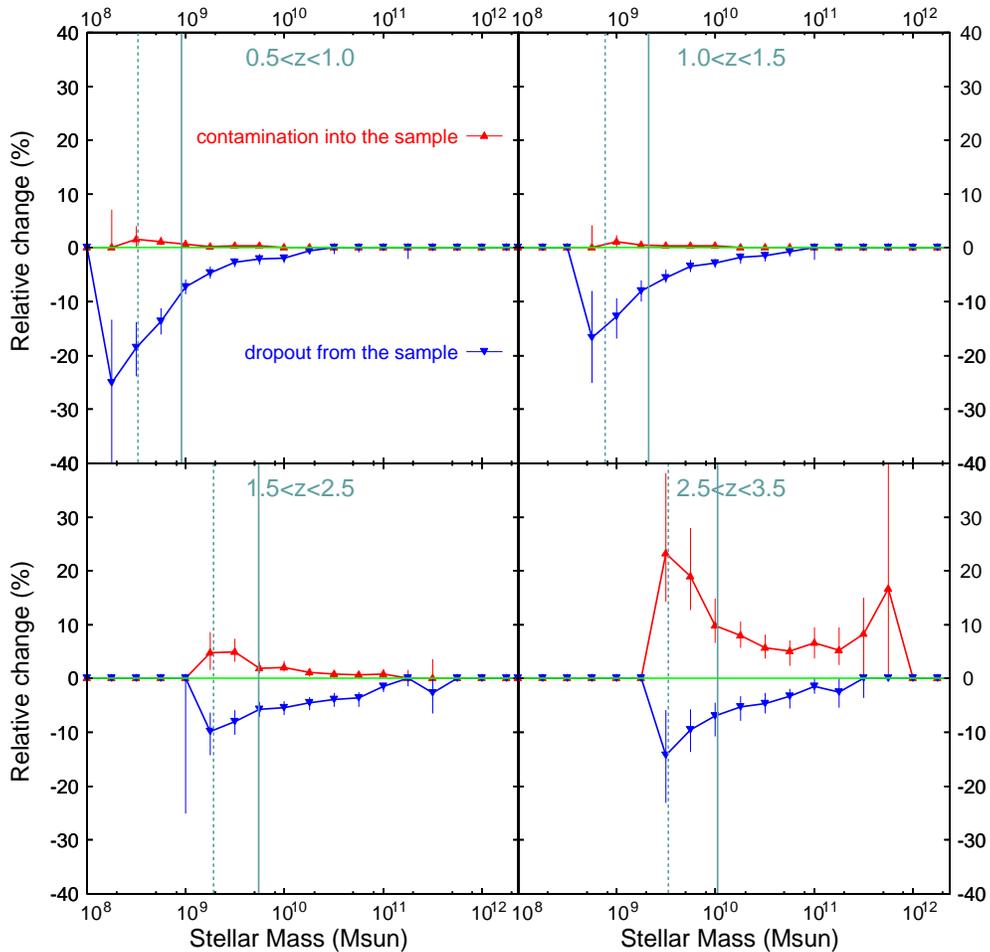}
\end{center}
\caption{Effect of the catastrophic failure of the 
photometric redshift on the SMF in the Monte Carlo simulation.
Upward and downward triangles show the fractional increase and decrease of 
the number of galaxies as a function of stellar mass 
in each redshift bin due to the objects 
whose photometric redshift was changed catastrophically 
($\delta z/(1+z) > 0.5$) by the random offsets of the multi-band photometry.
Upward (downward) triangles represent the fraction of the objects 
which enter into (drop out from) 
the redshift bin due to the catastrophic failure.
Data points and error bars represent the median value and 68 percentile   
interval of the 200 simulations.
Vertical lines show the limiting stellar mass  
for the wide (solid) and deep (dashed) samples.
}
\label{fig:catas}
\end{figure*}
\begin{figure}[ht]
\begin{center}
\includegraphics[angle=0,scale=0.6]{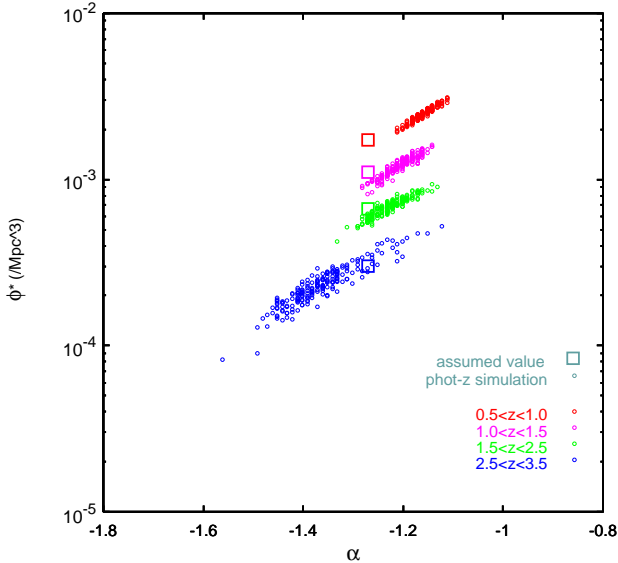}
\end{center}
\caption{Effect of photometric redshift uncertainty on
the normalization ($\phi^*$) of stellar mass function.
Large squares and small circles are the same as those in Figure \ref{fig:masim},
 but different colors represent different redshift bins.
The normalization is assumed to evolve with redshift as $\phi^{*}(z) \propto 
(1+z)^{-2}$ (see text).
}
\label{fig:phiasim}
\end{figure}
\begin{figure}[ht]
\begin{center}
\includegraphics[angle=0,scale=0.55]{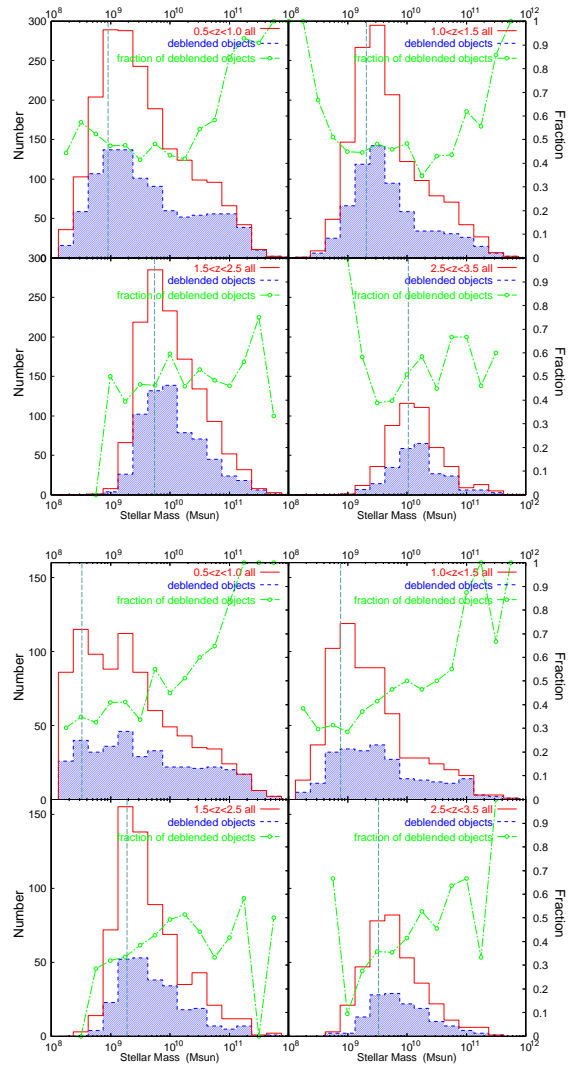}
\end{center}
\caption{Fraction of deblended objects as a function of stellar mass in each 
redshift bin for the wide (top) and deep (bottom) samples.
Solid line shows all $K$-selected galaxies in each redshift bin and 
shaded histogram represent deblended objects.
Dashed-dot line shows the fraction of the deblended objects.
Vertical long-dashed line shows the limiting stellar mass.
The results with the GALAXEV model are shown.
}
\label{fig:deblend}
\end{figure}
\begin{figure}[ht]
\begin{center}
\includegraphics[angle=0,scale=0.55]{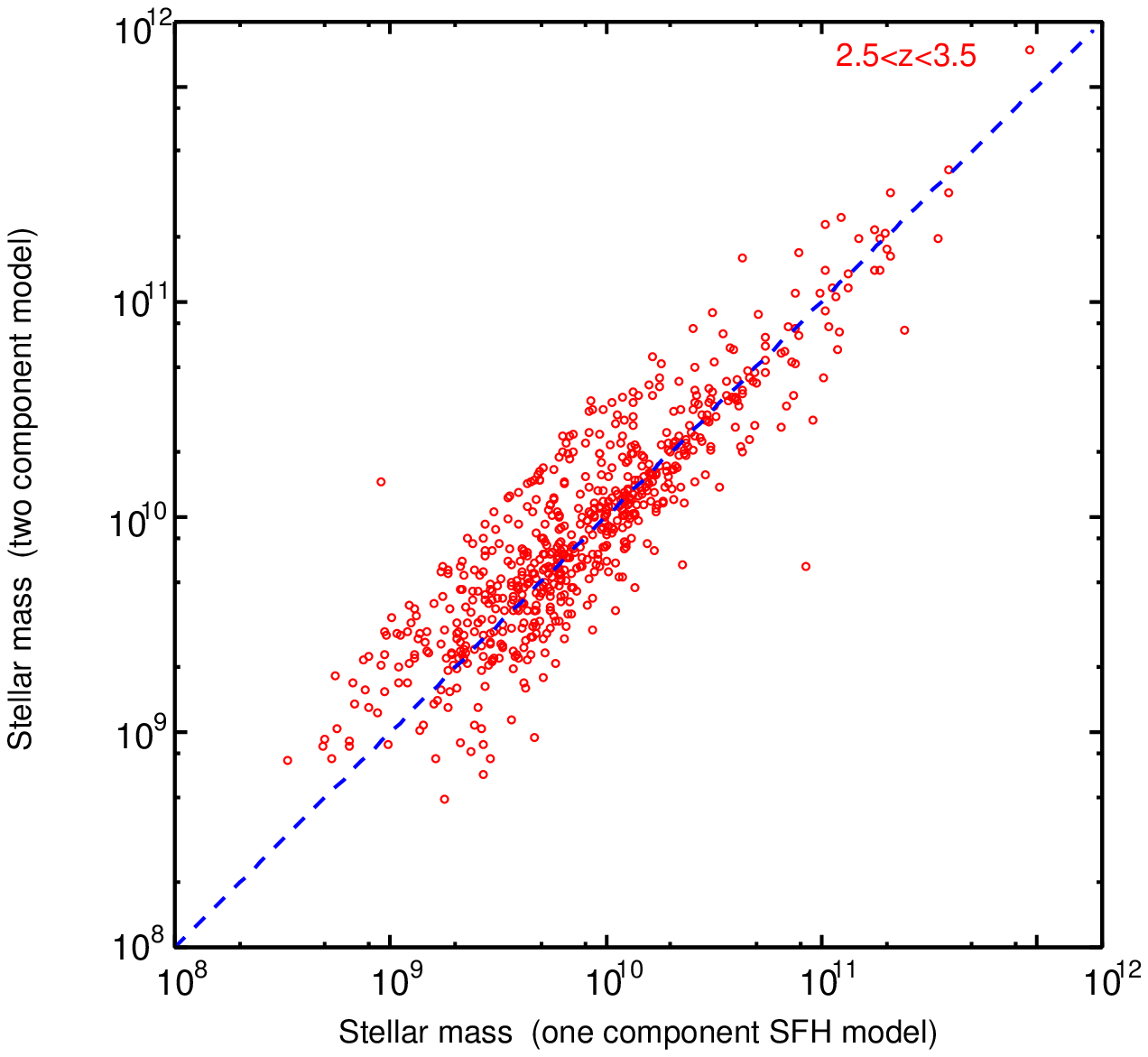}
\end{center}
\caption{Comparison between the stellar masses estimated with 
one-component SFH model and two-component (old and young) model
for galaxies at $2.5<z<3.5$. The result with the GALAXEV model is shown.
}
\label{fig:twocomp}
\end{figure}

We can see general evolutionary features  
in Figures \ref{fig:mfeach} and \ref{fig:mfall}
in spite of the field variance and the systematic differences 
among the SED models mentioned above.
We note, first, that the overall number density decreases gradually with 
redshift in all cases. While the number density of galaxies at $0.5<z<1.0$ 
is similar with that in the local universe, the number density at $2.5<z<3.5$ 
is about an order of magnitude smaller than the local value.
Figure \ref{fig:schpara} and Table \ref{tab:schpara} show the best-fit 
Schechter parameters ($\alpha$, M$^*$, $\phi^{*}$) estimated with the 
STY method. The redshift evolution of the overall number density 
can be seen as the decrease of the normalization of the SMF $\phi^{*}$.
$\phi^{*}$ decreases down to $\sim$ 50\% of that in the local universe 
at $z\sim0.75$, 
$\sim$ 16\% at $z\sim1.25$, $\sim$ 9\% at $z\sim2$ and 
$\sim$ 3\% at $z\sim3$.
Similar evolution of the SMF is also seen in 
previous studies of general fields (e.g., \citealp{fon06}, 
\citealp{per08}, \citealp{mar08}).

Second, we found the mass-dependent evolution of the SMF.
The evolution of the number density of 
low-mass galaxies with 
M$_{\rm star}\sim10^{9}$-10$^{10}$M$_{\odot}$ is 
smaller than that of massive galaxies 
with stellar mass $\sim10^{11}$M$_{\odot}$. While the number density
of galaxies with 
$\sim10^{11}$M$_{\odot}$ at $2.5<z<3.5$  
is smaller by a factor of $\sim$ 20 
than the local value, that of galaxies with 
$\sim5\times10^{9}$M$_{\odot}$ is 
smaller by only a factor of $\sim$ 6 at the same 
redshift. The shape of the SMF at $0.5<z<1.0$
is similar with that in the local universe, 
and it becomes steeper with redshift at $z>1$.
This can be seen in Figure \ref{fig:schpara} and Table \ref{tab:schpara} 
as a steepening of the low-mass slope $\alpha$ with redshift.
$\alpha$ decreases with redshift gradually 
from $\alpha = -1.29\pm0.03$($\pm0.04$)
at $0.5<z<1.0$ to $\alpha = -1.48\pm0.06$($\pm0.07$) at $1.5<z<2.5$ 
and $\alpha = -1.62\pm0.14$($\pm0.06$) at $2.5<z<3.5$.
The quoted errors are statistical errors estimated from the maximum likelihood
method. The values in parenthesis show the uncertainty due to the 
different SED models.

The uncertainty of the Schechter parameters becomes larger with redshift, 
especially at $2.5<z<3.5$, because of 
the larger limiting mass at higher redshift
and of the small number of galaxies even at relatively high mass 
due to the evolution of the overall number density mentioned above.
Nonetheless, since 
the uncertainty at $z<2.5$ is rather small by virtue of our deep and wide 
NIR data, the evolution of the low-mass slope of the SMF between  
$0\lesssim z \lesssim 3$ is found to be significant. 
Figure \ref{fig:malpha} shows the best-fit Schechter parameters and their 
uncertainty 
for the different SED models
in the M$^*$-$\alpha$ plane, which represents the evolution of 
the shape of the SMF.
In all cases, the evolution of $\alpha$ is significant, 
although there is degeneracy between M$*$ and $\alpha$ especially 
at $2.5<z<3.5$, where we can reach only to the relatively high stellar mass.

On the other hand, the characteristic mass M$^*$ shows no significant
evolution except for the results with the Maraston model.
The M$^*$ values at $0.5<z<3.5$ are similar with or slightly larger than those
in the local universe. No significant evolution 
for M$^*$ 
is also seen in previous 
 studies (\citealp{fon06}, \citealp{poz07}, \citealp{mar08}).  
For the 
Maraston model, M$^*$ becomes smaller by a factor of $\sim$ 2--2.5 at $z>1.5$.
As in the above discussion of the overall number density, this can be explained
by the systematically lower stellar M/L ratio of the Maraston model because 
the same result is also seen when the same photometric redshifts (EAZY)
are used. While passively evolving galaxies dominate the massive end 
($\gtrsim10^{11}$M$_{\odot}$) of the 
SMF at $z\lesssim1$ (e.g., \citealp{jun05}, \citealp{bor06}, \citealp{ver08}, 
\citealp{ilb09}), many massive 
star-forming (i.e., relatively young) galaxies have been found at $z\sim2$ 
(e.g., \citealp{dad07}, \citealp{pap06}, \citealp{bor05}, \citealp{sha04}).
It is possible that TP-AGB stars contribute to the SED of massive 
galaxies significantly only at $z>1.5$.
Therefore, the Maraston model would give systematically 
lower stellar mass for these massive galaxies.
\begin{figure*}[t]
\begin{center}
\includegraphics[angle=0,scale=0.9]{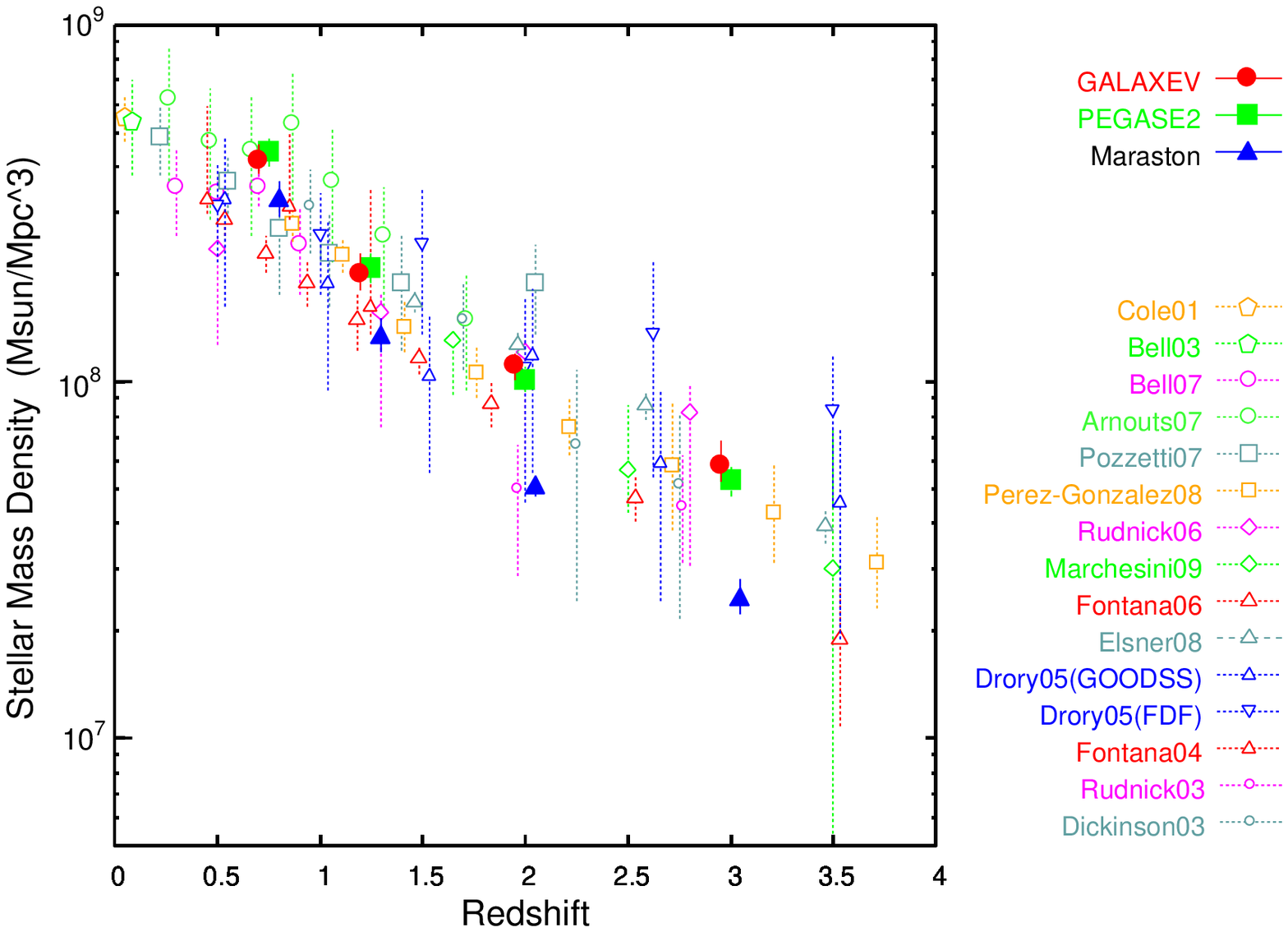}
\end{center}
\caption{Evolution of the stellar mass density 
(integrated over 10$^8$M$_{\odot}<$M$_{\rm star}<10^{13}$M$_{\odot}$) 
as a function of redshift for the different 
SED models (solid symbols).
Error bars are based on the Poisson statistics.
 Open symbols show 
results from previous surveys (compilations from \citealp{wil08} and 
\citealp{mar08}). 
Some data points are shifted horizontally for 
clarity. 
}
\label{fig:msd}
\end{figure*}
\begin{figure*}[t]
\begin{center}
\includegraphics[angle=0,scale=1.0]{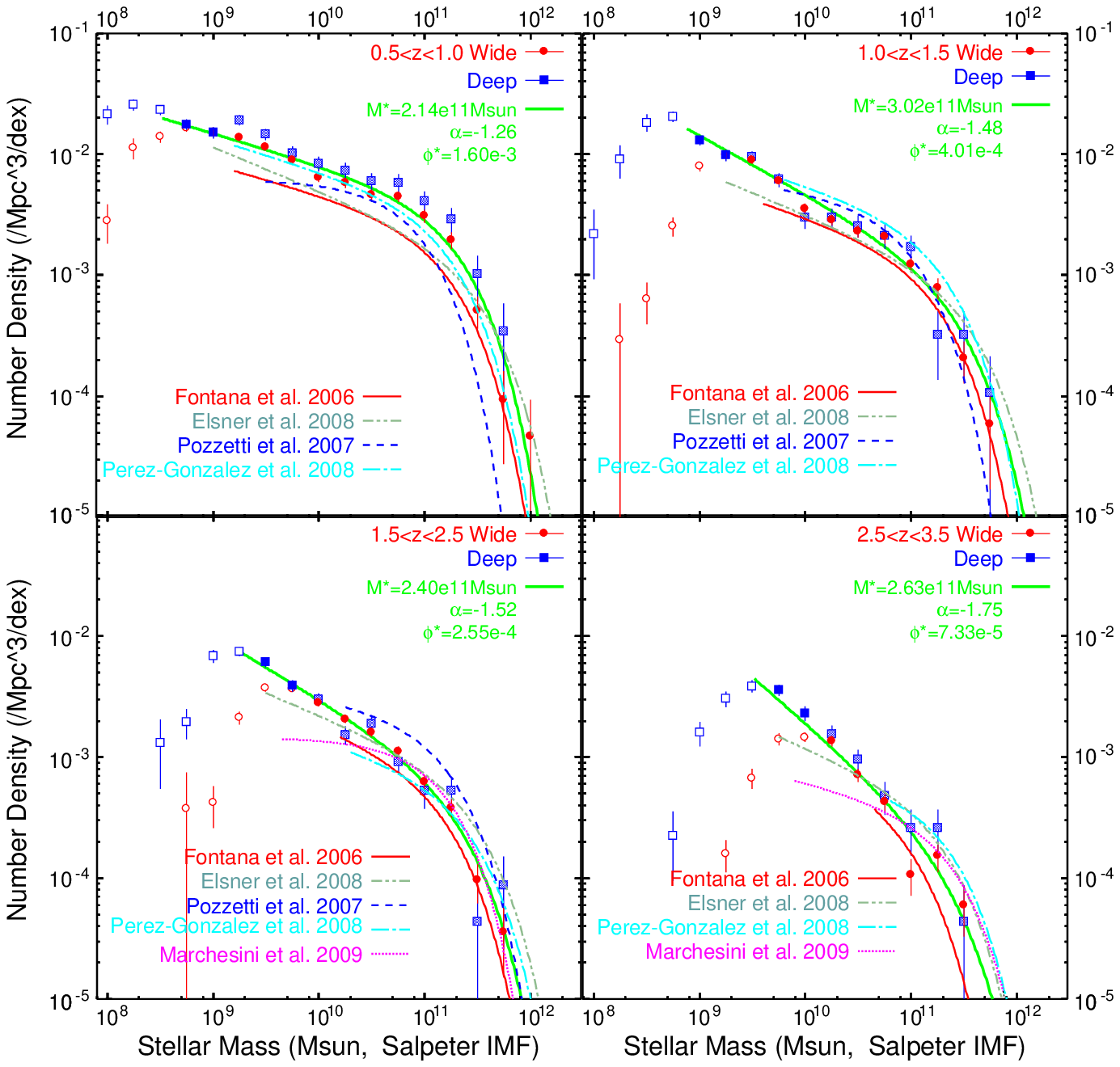}
\end{center}
\caption{Comparison between stellar mass functions in this study and  
 other surveys. The GALAXEV model 
is used for the direct comparison with previous studies.
Circles, squares and solid line are the same as those in 
Figure \ref{fig:mfeach}. 
For other deep surveys, 
the best-fit Schechter functions from the literature (\citealp{fon06}, 
\citealp{els08}, \citealp{poz07}, \citealp{per08}, \citealp{mar08}) are plotted over 
 the observed mass range.
}
\label{fig:mscomp}
\end{figure*}

\subsection{Possible biases for the evolution of the low-mass slope}

Here we investigate possible biases for the results in the previous subsection,
in particular, systematic effects which could cause the steepening 
of the low-mass slope at high redshift. 
 
The larger limiting mass at higher redshift causes the degeneracy between
$M^*$ and $\alpha$ as mentioned in the previous section. 
Since low-mass galaxies near the limiting mass 
tend to be faint in each band, 
the photometric errors are larger, which results in 
the large uncertainty in
 their photometric redshift. The large errors of the 
photometric redshifts of faint objects might lead to the systematic increase 
of the low-mass galaxies at high redshift because the redshift distribution 
of the $K$-selected sample has a peak around $z\sim1$ and 
a tail to higher redshift. 
In order to evaluate the effect on the low-mass slope, we performed 
a Monte Carlo simulation, assuming no evolution of the shape of the SMF (i.e., 
constant M$^*$ and $\alpha$). 
At first, we constructed mock catalogs with 
the observed M$^*$ and $\alpha$ at $0.5<z<1.0$. 
For the normalization of the SMF, we assumed $\phi^{*}$ evolving  
as $\phi^{*}(z) \propto (1+z)^{-2}$ so that the redshift distribution of 
the mock sample is consistent with our observation.
Even if we assume $\phi^{*}(z) \propto (1+z)^{-1}$ or $\phi^{*}(z) \propto (1+z)^{-3}$, the results shown in the following do not change significantly.
Stellar mass and redshift of mock objects   
were randomly selected from the ranges of 
$10^{8}$ M$_{\odot}<$ M$_{\rm star}<10^{12}$ M$_{\odot}$ and $0<z<6$, 
using the probability distribution estimated from 
the assumed SMF and the corresponding comoving volume of our survey 
at each redshift. 
For each mock object, we randomly extract an object from the 
observed sample with similar mass and redshift (allowing duplicate), 
and adopted its observed multi-band photometry. 
The observed multi-band photometry was extracted from the catalog that contains 
all sources detected on the $Ks$-band image (including $K>23$ 
or $K>24$ objects) in order to
take into account the scattering of objects fainter than the magnitude limits. 
Then we added random offsets to the multi-band photometry 
according to the measured photometric errors, and adopted the mock object 
if the resulting $K$-band magnitude of the object was brighter than the 
magnitude limits ($K<23$ for the wide sample and $K<24$ for the deep sample).
We repeated this procedure and made the mock catalogs with the same 
sample sizes as the observed wide and deep samples.
The same SED fitting procedure as for the observed one was performed in order 
to estimate the photometric redshift and stellar mass of the mock objects. 
For objects with 
spectroscopic identification, redshifts are fixed to the spectroscopic values.
We performed 200 simulations and calculated the best-fit Schechter parameters 
in each simulation with the STY method.

Figure \ref{fig:masim} shows the results of the simulation in the case with 
GALAXEV (the results for the other models are similar).
A relatively large scatter of $\alpha$ is seen in the highest redshift bin, 
which probably reflects the large 
degeneracy between M$^{*}$ and $\alpha$ due to 
the large limiting mass. 
Furthermore, the simulated $\alpha$ distributes around 
a systematically steeper (by $\sim$ 0.1--0.15) value than the assumed one at 
$2.5<z<3.5$, 
while the simulated values tend to be slightly 
flatter (by $\sim$ 0.1) in lower redshift bins.
However, since the observed evolution of $\alpha$ is much stronger than the 
systematic effects in Figure \ref{fig:masim}, the observed 
steepening of the low-mass 
slope at high redshift is significant, especially at $1<z<2.5$.

In the simulation, the random offsets added to the multi-band photometry 
and the recalculation of the photometric redshift include the effect of 
the catastrophic failure. We extracted the mock objects whose photometric 
redshift was changed catastrophically
($\delta z/(1+z) > 0.5$) by the random offsets 
 and checked the effect of these objects on the resulting SMF.
Figure \ref{fig:catas} shows the fractional increase and decrease of the number of 
galaxies due to the catastrophic failure of the photometric redshift 
as a function of stellar mass in each redshift bin. 
At $0.5<z<1.5$, 
there is $\sim$ 10-20\% decrease near the limiting stellar mass, 
while there is only negligible effect of the catastrophic failure at 
M$_{\rm star} \gtrsim 10^{10}$M$_{\odot}$. Most contamination from $z<0.5$ or 
$z>1.5$ occurs only at the stellar mass lower than limiting mass and 
it is not plotted in the figure. 
About 10-20\% decrease near the limiting stellar mass might cause 
a slightly flatter low-mass slope seen in Figure \ref{fig:masim}, 
but the effect 
is relatively small ($\sim$0.1 dex decrease in the number density).
At $z>1.5$, the effect of the catastrophic failure is similar or even smaller 
than that at low redshift. Furthermore, the contamination from lower redshift 
and the dropout from the redshift bin tend to be canceled out with each other, 
which results in the negligible effect on the SMF. 

In Figure \ref{fig:phiasim}, we also show the results of the same simulation 
on the $\alpha$-$\phi^{*}$ plane to investigate the effect on the normalization.
The systematic effect on the normalization of the SMF due to the photometric 
redshift errors seems to be relatively small, although the degeneracy 
between the parameters makes direct comparison difficult.

Next, we discuss a possibility of the over-deblending of faint objects with 
relatively low S/N ratio near the detection limit. Although the $K$-band,  
 where we performed the source detection, corresponds to the rest-frame $B$-band
even at $z=3.5$, the morphological K-correction could enhance the 
over-deblending at high redshift because galaxies  
tend to show the patchy 
appearance in shorter wavelengths due to the dominance of young stars and the 
dust extinction (e.g., \citealp{kuc01}, \citealp{raw09}).
Figure \ref{fig:deblend} shows the fraction of the objects with the deblending 
flag by SExtractor as a function of stellar mass in each redshift bin for 
the wide and deep samples. We cannot see the mass nor 
 redshift dependence of the fraction of the deblended objects.
We conclude that 
the effect of deblending does not cause the evolution of the low-mass slope.

Finally, we estimated the stellar mass of galaxies in the highest redshift bin  
with the GALAXEV templates 
with two-component (old and young) star formation histories.
If the old stellar population is hidden by 
 recent star formation, the stellar mass could be underestimated 
especially for relatively blue low-mass galaxies at high redshift, 
where the S/N ratio tends to be low (e.g., \citealp{pap01}, \citealp{dro05}).
We used the exponentially decaying star formation models (young component) with 
an old population component. For the young population, free parameters are 
stellar age, star formation timescale $\tau$, color excess, metallicity 
(same as for one-component star-formation history).
For the old component, we limited the star formation timescale and stellar 
age to shorter and older values than the young population, respectively, and 
assumed no dust extinction. Figure \ref{fig:twocomp} shows 
the comparison of the 
stellar masses estimated with one- and two-component models 
for galaxies at $2.5<z<3.5$. 
No significant systematic difference
of the stellar mass can be seen. 
The scatter is consistent with the uncertainty 
of the stellar mass of these galaxies shown in Figure \ref{fig:sigms}, 
although the stellar mass with the two-component model is slightly 
larger for a small fraction 
of galaxies at low-mass end. 
The long wavelength data with Spitzer/IRAC, which 
sample the rest-frame NIR region even at high redshift, 
could make the systematic 
uncertainty relatively small (\citealp{fon06}, \citealp{els08}).
\begin{figure*}[t]
\begin{center}
\includegraphics[angle=0,scale=1.1]{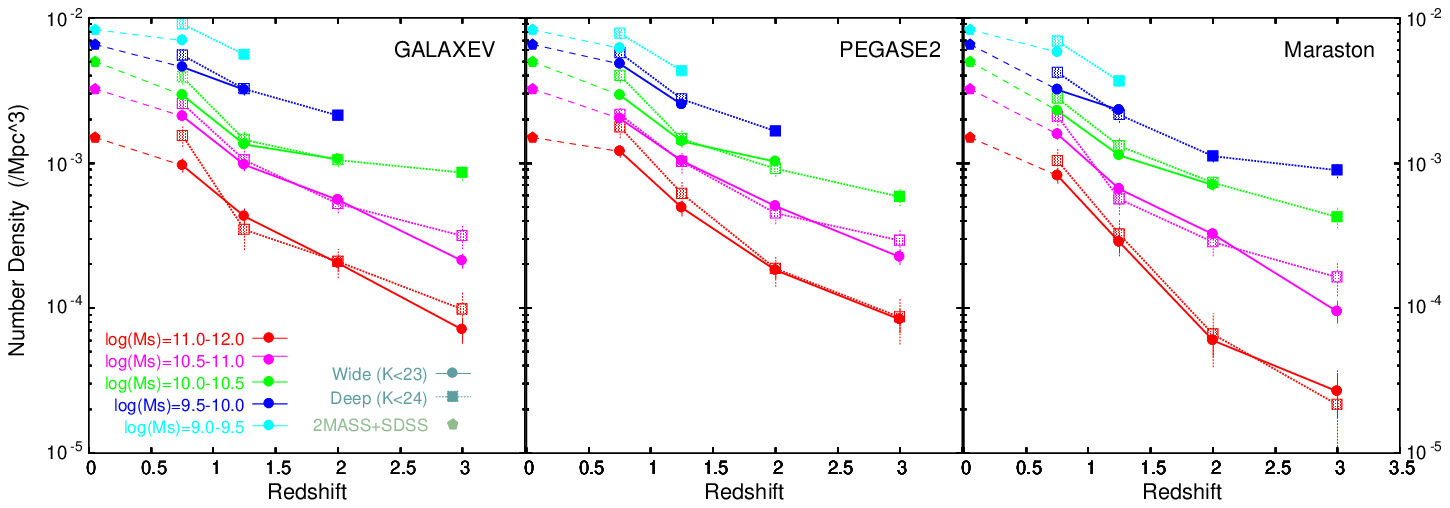} 
\end{center}
\caption{Evolution of the number density of galaxies in different stellar mass  
ranges. Three panels represent the results with the different SED models. 
Circles and squares show the wide and deep samples, respectively.
Different colors of the 
symbols represent different ranges of the stellar mass of 
galaxies. 
Only stellar mass ranges which are above the limiting stellar mass 
in Figure \ref{fig:mlim} at each redshift are shown. 
}
\label{fig:ndeach}
\end{figure*}
\begin{figure*}[ht]
\begin{center}
\includegraphics[angle=0,scale=1.1]{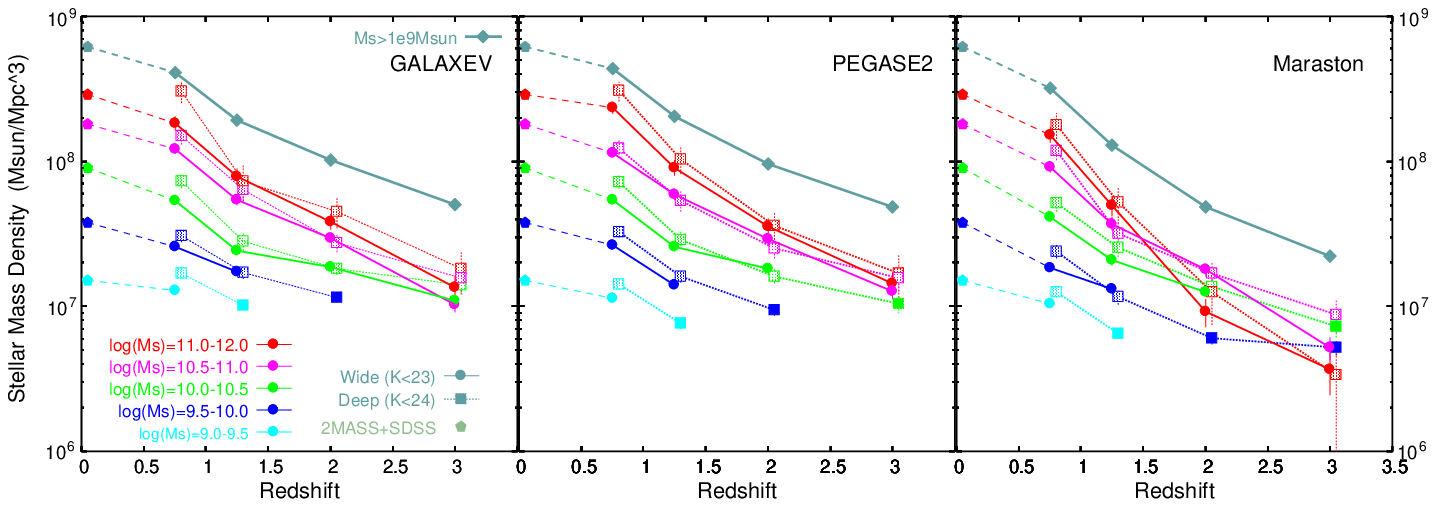}
\end{center}
\caption{Contributions of different stellar mass ranges
to the total stellar mass density as a function of redshift.
Symbols are the same as in Figure \ref{fig:ndeach}.
Diamonds show the stellar mass density integrated 
over $10^{9}$M$_{\odot}<$M$_{\rm star}<10^{12}$M$_{\odot}$ at each redshift.
}
\label{fig:mdeach}
\end{figure*}

\section{Discussion}

We have studied the evolution of the SMF of galaxies back to 
$z\sim3$ using the NIR Subaru/MOIRCS data obtained in the MOIRCS Deep Survey
and the publicly available multi-wavelength GOODS data.
Our deep and wide NIR data allowed us to estimate 
the number density of galaxies down to 
low stellar mass ($\sim10^{9}$-10$^{10}$M$_{\odot}$) even at $z\sim$ 2--3
with high accuracy.
Main results in the previous section are 1) the decrease of the overall number
density of galaxies with redshift, and 2) the steepening of the low-mass slope
of the SMF at high redshift. 
In this section, we first compare the results with previous studies 
and then discuss the implications for 
galaxy formation and evolution in the following. 
\begin{figure*}[t]
\begin{center}
\includegraphics[angle=0,scale=1.0]{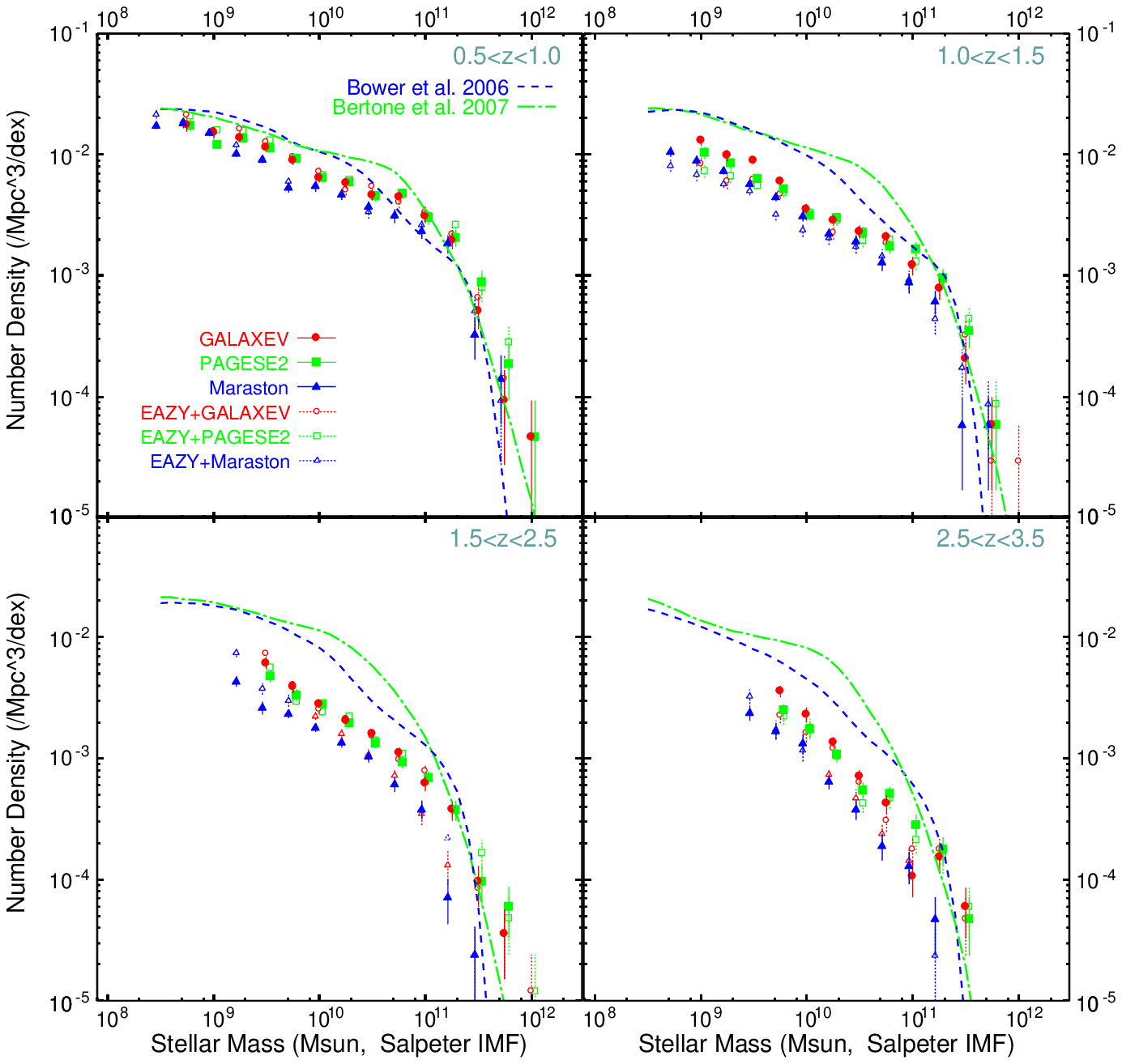}
\end{center}
\caption{Comparison of the observed stellar mass function 
with the predictions by the semi-analytic models.
Short-dashed line refers to the \citet{bow06} model and dashed-dot line 
represents the \citet{ber07} model.
Circles, squares and triangles are the same as in Figure \ref{fig:mfall}. 
}
\label{fig:sma}
\end{figure*}

\subsection{Comparison with previous studies}

In this subsection, we compare our results in the MODS field with those in 
other general fields.

Figure \ref{fig:msd} shows a comparison of the integrated 
stellar mass density as a function of redshift. 
We calculated the stellar mass density 
by integrating the best-fit Schechter function over 10$^{8}$-10$^{13}$ 
M$_{\odot}$ in each redshift bin. Different symbols represent the results with 
the different SED models. Compilations of  
\citet{wil08} and \citet{mar08} are also plotted.  
All data points are the results with the Salpeter IMF. 
The stellar mass density in the MODS field decreases with redshift
as seen in other previous studies. 
When compared with those in the local universe, 
the stellar mass density is $\sim$ 53--72\% of the local value at $0.5<z<1.0$, 
$\sim$ 22--34\% at $1.0<z<1.5$, $\sim$ 8--18\% at $1.5<z<2.5$ and 
$\sim$ 4--9\% at $2.5<z<3.5$.
For the GALAXEV model, which is widely used in other
previous studies (e.g., \citealp{dro05}, 
\citealp{fon06}, \citealp{poz07}, \citealp{els08}, \citealp{mar08}), 
our results are consistent with those of previous 
studies at each redshift (Figure \ref{fig:msd}). 
A slightly large value at $0.5<z<1.0$ is 
probably due to the large scale structures around the HDF-N mentioned above.
As discussed in Section \ref{sec:result}, the Maraston model gives 
systematically lower stellar mass density, which is about three-quarters of 
those with the other two models (GALAXEV and PEGASE2) at $0.5<z<1.0$, 
about two-thirds at $1.0<z<1.5$ and about a half at $z>1.5$.

In Figure \ref{fig:mscomp} we compare the SMF 
of galaxies in the MODS 
field with those in other studies. For simplicity, we plotted the result with 
the GALAXEV model.
At the massive end ($\gtrsim 10^{11}$M$_{\odot}$), our result is consistent 
with those in other general fields, although the uncertainty of our data and 
the variance among the different surveys are relatively large because of 
the small number of observed massive galaxies and probably 
their strong clustering (e.g., \citealp{vdo06}, \citealp{con07}, \citealp{str09}). 
The agreement of the SMF at the massive end is consistent with the 
comparison of the integrated stellar mass density because 
galaxies around M$^*$ ($\sim10^{11}$M$_{\odot}$) dominate the stellar mass 
density (e.g., \citealp{gwy05}, see also next subsection).
No significant evolution of M$^*$ seen in the MODS field 
at $0.5<z<3.5$ is also consistent 
with results of previous studies (\citealp{fon06}, \citealp{ber07}, 
\citealp{poz07}, \citealp{els08}, \citealp{per08}, \citealp{mar08}). 

On the other hand, at the lower mass of the SMF, 
there is some variance in different fields 
including the MODS field in spite of the relatively small statistical errors.
In addition to the differences in the normalization, which are similar offsets  
with those at the massive end, there is  
differences in the low-mass slope especially at high redshift. 
 One possible reason for this is the different depths of the surveys and 
therefore the different limiting stellar masses. 
The SMF in the MODS field 
shows a slight upturn around $\sim$10$^{10}$M$_{\odot}$ and a steeper slope 
at $<10^{10}$M$_{\odot}$, although this is marginal in the highest redshift
bin. This indication can be seen in the results with all SED models (Figure 
\ref{fig:mfeach}) and does not depend on the systematic error of 
the photometric redshift and stellar M/L ratio. Similar upturns 
in the SMF are also seen in field galaxies in the local universe \citep{bal08}
 and  at intermediate and high redshift (\citealp{fon06}, \citealp{poz07}, 
\citealp{els08}). 
Therefore the relatively shallow survey with the 
limiting stellar mass larger than $\sim10^{10}$M$_{\odot}$ could miss 
 the upturn and 
 result in the systematically flatter low-mass slope even if 
the completeness to the limit is sufficiently high or the incompleteness 
correction is properly done. 
For example, the SMFs from the different surveys show consistent low-mass 
slopes in the $0.5<z<1.0$ bin in Figure \ref{fig:mscomp}, 
where all the surveys reach below 10$^{10}$M$_{\odot}$. 
At $1.0<z<1.5$, 
several surveys reach only to $\sim$10$^{10}$M$_{\odot}$.
While the slope at 10$^{10}$-10$^{11}$M$_{\odot}$
 is similar with those in other surveys, 
the SMF in the MODS field shows the steeper slope at 10$^9$-10$^{10}$
M$_{\odot}$ than that at larger stellar mass.
This could be the case in the higher redshift bins.

At $z>1.5$, on the other hand, 
\citet{els08} and \citet{mar08} reach below 10$^{10}$M$_{\odot}$ 
and show the flatter low-mass slopes than those in the present study.
 In \citet{els08}, the low-mass slope is fixed to the weighted mean 
value $\alpha\sim -1.36$, which is constrained mainly by galaxies at 
relatively low redshift ($z\lesssim1$, their Figure 7).
\citet{mar08} discussed the effects of various random 
and systematic uncertainties on the estimate of the Schechter parameters of 
the SMF and suggested that the uncertainty of the low-mass slope 
in their analysis 
is relatively large especially at high redshift due to the statistical error 
and the systematic errors of the different SED models. 
The difference in the low-mass slope 
between \citet{mar08} and the present study could be attributed to  
field-to-field variance. 
In the local universe, several studies suggest that  
the shape of the SMF depends on the environment of galaxies (\citealp{bal01},
 \citealp{bal06}, \citealp{bal08}), and this could be the case at higher redshift
(e.g., \citealp{sco09}). 
In \citet{mar08}, the low-mass slope at high redshift 
is constrained mainly by the FIRES data, whose survey area is about a fifth 
of our study. 
If the shape of the SMF depends on the environment even at 
high redshift, the low-mass slope estimated from the small area surveys 
could be affected by the field-to-field variance. 
A larger area survey with a similar or fainter limiting magnitude in the NIR 
wavelength is needed to 
conclude whether there exists the field-to-field variance of the shape of 
the SMF at $z\sim$ 2--3 or not. 

\subsection{Mass-dependent evolution of galaxies}

The decrease of the overall number density and the steepening of the 
low-mass slope with redshift seen in the evolution of the SMF of galaxies 
in the MODS field suggest the possibility of 
the mass-dependent number density evolution.
Figure \ref{fig:ndeach} shows the evolution of the number density of 
galaxies in 
different stellar mass ranges for the different SED models. 
Relatively low-mass galaxies with the stellar mass of 
10$^{9.5}$-10$^{10.5}$M$_{\odot}$ tend to 
show a weaker evolution of the number density 
 especially at $z\gtrsim1.5$ than M$^{*}$ ($\sim10^{11}$M$_{\odot}$) galaxies, 
although the trend is relatively marginal in the results with the 
GALAXEV/PEGASE2 models.
The relative number density 
of low-mass galaxies had been larger at high redshift than 
that in the local universe. 
In other words, the number density of galaxies around M$^{*}$ 
might have more rapidly 
increased at $z\gtrsim1.5$ than that of low-mass galaxies.
 On the other hand, several studies based on 
 wide area but shallow surveys suggest that for M$_{\rm star} > 
10^{11}$M$_{\odot}$, more massive (e.g., $\gtrsim$ 10$^{11.5}$-10$^{12}$M$_{\odot}$)
 galaxies show weaker evolution in the number density at $z\gtrsim1.5$ 
(\citealp{brt07}, \citealp{per08}). 
Although the survey volume of the present study is not large enough  
to constrain strongly the number density of very massive galaxies with 
10$^{11.5}$-10$^{12}$M$_{\odot}$, galaxies around $10^{11}$M$_{\odot}$ 
might have been formed effectively at $1\lesssim z \lesssim 3$ compared with 
higher and lower mass galaxies (e.g., \citealp{fon04}, \citealp{fra06}).

Figure \ref{fig:mdeach} shows the contribution of galaxies in different 
stellar mass ranges to the cosmic stellar mass density. 
While the stellar mass density is dominated 
by massive galaxies with M$_{\rm star}>10^{10.5}$M$_{\odot}$ 
in the local universe, 
the contribution of lower mass galaxies increases with redshift and is 
 significant at $z\sim3$. 
\citet{red09} reported a relatively steep faint-end slope ($\alpha \sim -1.7$)
of the luminosity function for Lyman Break Galaxies (LBGs) at $1.9<z<3.4$.
They estimated the stellar mass of faint 
LBGs with the log-linear relation between SFR and stellar mass 
derived in \citet{saw07}, and suggested that those faint LBGs, which typically
have M$_{\rm star} \lesssim 10^{10}$M$_{\odot}$ (Figure 12 in \citealp{red09}), 
show a significant contribution to the total stellar mass density.
Since most low-mass galaxies at $z\sim$ 2--3 have rather blue rest-frame colors
(typically rest $U-V\lesssim$0, Figure \ref{fig:uvms}, \citealp{kaj05}, 
\citealp{kaj06b}) and satisfy the LBG criteria \citep{lab05}, 
the steep low-mass 
slope of the SMF at high redshift in our study seems to be consistent with 
the result in \citet{red09}. Note that   
the integrated stellar mass density 
in the MODS field is consistent with 
 those of other previous studies as mentioned 
in the previous subsection, but about a half of the estimate in \citet{red09}. 
Then, although the contribution of low-mass 
galaxies to the total stellar mass density 
in the MODS field is larger than those estimated in previous studies, 
the stellar mass density is still smaller than 
that expected from the evolution of the cosmic SFR density 
(e.g., \citealp{wil08}, \citealp{red09}).

\subsection{The origin of the evolution of the low-mass slope}

What is the origin of the evolution of the low-mass slope ?
In hierarchical structure formation scenarios, low-mass objects continue 
to grow by merging over cosmic time and result in more massive objects at later
epochs. Therefore one can expect the steeper 
low-mass slope at the higher redshift (e.g., \citealp{kho07}, 
\citealp{rya07}). Figure \ref{fig:sma} compares the observed SMF in the MODS 
field with the predictions of the publicly available 
semi-analytic models by \citet{bow06} and 
\citet{ber07}. These two models are based on the Millennium Simulation of 
the growth of dark matter structure in a $\Lambda$CDM cosmology \citep{spr05} 
but are made by the different procedures in 
many aspects such as, e.g., the construction of halo merger trees, 
the implementation of the star formation, 
the feedback by supernova explosions and AGN activities.
We convolved these model predictions with the Gaussian representing the 
uncertainty of the stellar mass estimate shown in Figure \ref{fig:sigms}. 
The median value in Figure \ref{fig:sigms} is chosen 
as the width of the Gaussian depending on galaxy's stellar mass and redshift.
In the models, the low-mass slope similarly steepens 
 with redshift in Figure \ref{fig:sma}. 
However, it should be noted that there are discrepancies between 
the observation and the models in the detailed shape of the SMF and that 
the evolution of the overall number density of model galaxies 
is significantly weaker 
(i.e., larger number density at high redshift)
than that of the present observation  
especially at low mass as in previous studies (e.g., \citealp{fon06}, 
\citealp{kit07}, \citealp{ber07}). 
On the other hand, several studies  
found that semi-analytic models tend to predict the smaller 
number density of galaxies at the massive end
(M$_{\rm star}\gtrsim 10^{11.5}$-$10^{11.75}$M$_{\odot}$)
than observations (e.g., \citealp{con07}, \citealp{ber07}). 
A similar indication is also seen in Figure \ref{fig:sma},  
although our data cannot strongly constrain the number density of these very 
massive galaxies. 

Since the stellar mass of a galaxy is generally dominated by long-lived 
low-mass stars, it can be generally considered as the integral of the
past star formation rate and expected to increase monotonically with time. 
Therefore, for example,  only 
suppression of star formation in low-mass galaxies 
cannot cause the flattening of the low-mass slope of the SMF with time, 
while it could lead to the evolution of the faint-end slope of the rest-frame 
UV luminosity function, which directly reflects the star formation rate, 
as discussed in \citet{red09}.
On the other hand, 
the stellar mass growth by star formation activities 
in each galaxy could change the slope of the SMF. 
There are several observational studies that suggest that low-mass galaxies 
 have higher star formation rate relative to their stellar mass 
(i.e., SFR/M$_{\rm star}$, specific star formation rate) than massive galaxies
even at $z\sim$ 2--3 (e.g., \citealp{feu05}, \citealp{pap06}, \citealp{red06b}, 
see also \citealp{san09}). 
These results for  
the mass dependence of SFR/M$_{\rm star}$ suggest that 
the expected growth rate of stellar mass by star formation activities 
in a galaxy is higher in lower mass galaxies.
If we assume the observed trends of the low-mass slope and mass dependence of 
SFR/M$_{\rm star}$ continue down to lower mass, 
a net increase of the number of galaxies in a mass bin due to 
 star formation activities 
(inflow from lower mass bins $-$ outflow into higher mass bins) 
is expected to be larger at lower stellar mass, 
which leads to 
the steepening of the low-mass slope with time, 
i.e.,  opposite to that seen in Figures \ref{fig:mfeach} and 
\ref{fig:mfall}.  
For example, \citet{dro08} calculated the expected 
change of the SMF due to star formation only from the average star formation 
rate at each stellar mass of galaxies in the FORS Deep Field \citep{feu05}
and showed that the number density of lower mass galaxies is expected to
increase more rapidly with time especially at $z\lesssim$ 3. 
By subtracting the expected 
contribution due to star formation from the observed redshift evolution
 of the SMF, \citet{dro08} evaluated the effect of the hierarchical 
merging on the evolution of the SMF (i.e., destroying small galaxies and 
building more massive ones). 
In their analysis, at M$_{\rm star}\lesssim 10^{10}$M$_{\odot}$, 
lower mass galaxies tend to be consumed in mergers at higher rate 
to counterbalance the effect of the mass-dependent star formation rate.
Since no evolution of the low-mass end slope is assumed in \citet{dro08}, 
the steepening with redshift  
indicates the effect of merging larger than that estimated in \citet{dro08}.
Therefore the evolution of the SMF of galaxies in the MODS field suggests that 
the hierarchical merging process was important for 
the stellar mass assembly of these low-mass galaxies at $z\sim$ 1--3, when 
the cosmic stellar mass density had been increasing rapidly.
The relatively high fraction of peculiar/irregular morphology 
in low-mass galaxies at 
$z>1$ in the NIR wavelength (\citealp{kaj05}, \citealp{con05}) 
might also imply intense merging activity of these galaxies.

In this context, the fact that low-mass galaxies tend to have  
relatively younger stellar age and higher SFR/M$_{\rm star}$ than 
massive galaxies at low to intermediate redshift  
(i.e., `downsizing', e.g., \citealp{kau03}, \citealp{hea04}) might  
not necessarily mean relative scarcity of low-mass galaxies at high redshift.
A significant fraction of old stars in relatively 
massive (e.g., $\sim10^{11}$M$_{\odot}$) galaxies at low redshift 
could be formed at high redshift in actively star-forming low-mass galaxies 
which had coalesced into more massive objects through the mergers.
One of the means to address this is to investigate the distribution of 
the SFR and 
mean stellar age of galaxies as a function of stellar mass down to low mass 
at high redshift. The combination of the distribution of the SFR and 
mean stellar age as a function of stellar mass and the SMF shown in this study 
allows 
us to measure the contributions of galaxies in a certain mass range to the 
cosmic star formation rate density and stellar mass density and their 
age distribution as a function of redshift. 
We will be able to use these quantities in order to 
estimate what the mass range of high-z 
galaxies could have assembled into more massive galaxies at later epochs. 


\section{Summary}

In this paper, we have investigated the stellar mass function of galaxies 
at $0.5<z<3.5$, using the very deep and wide NIR imaging data obtained in 
MOIRCS Deep Survey and the multi-wavelength public data from the GOODS. 
The MODS data reach $K\sim23$ over $\sim$ 103 arcmin$^2$ and
 $K\sim24$ over $\sim$ 28 arcmin$^2$. 
We constructed a   
large sample of galaxies down to $\sim$ 10$^9$-10$^{10}$M$_{\odot}$ 
up to $z\sim$3.

Our main results are as follows:
\begin{itemize}
\item The normalization of the SMF decreases  
with redshift gradually 
and the integrated stellar mass density becomes $\sim$ 8--18\% of the local 
value at $z\sim2$ and $\sim$ 4--9\% at $z\sim3$. 
The evolution of the stellar mass density estimated with the GALAXEV model 
is consistent with 
those in previous surveys. 
\item The characteristic mass M$^*$ of the best-fit Schechter function shows  
no significant evolution. With the Maraston model, however, M$^*$ becomes 
smaller by a factor of $\sim$ 2--2.5 at $z>1.5$. This may be because 
 many massive galaxies are in active star-forming phase 
at $z>1.5$ and TP-AGB stars would contribute to the SED of these massive  
galaxies significantly only at $z>1.5$. 
\item The low-mass slope of the SMF becomes steeper with redshift 
gradually 
from $\alpha = -1.29\pm0.03$($\pm0.04$)
at $0.5<z<1.0$ to $\alpha = -1.48\pm0.06$($\pm0.07$) at $1.5<z<2.5$ 
and $\alpha = -1.62\pm0.14$($\pm0.06$) at $2.5<z<3.5$. 
The evolution of the number density of low-mass 
(10$^{9}$-10$^{10}$M$_{\odot}$) galaxies is weaker than that of 
M$^{*}$ ($\sim10^{11}$M$_{\odot}$) galaxies. The contribution of these low-mass galaxies 
to the cosmic stellar mass density increases with redshift, and becomes  
significant at $z\sim3$. 
\item We noted a marginal upturn around $\sim10^{10}$M$_{\odot}$ 
and a steeper slope at $<10^{10}$M$_{\odot}$ in the SMF. 
Shallow data, which does not reach to $<10^{10}$M$_{\odot}$,  
could lead to underestimation of the low-mass slope. 
\item The steepening of the low-mass slope with redshift
could be explained as a result of 
the hierarchical merging process. 
Since the mass dependence of the SFR/M$_{\rm star}$ 
distribution seen in previous studies 
is expected to lead to the steepening of the low-mass slope with time, 
the opposite trend found in this study suggests that 
the hierarchical merging process was very important for the stellar mass 
assembly of relatively low-mass galaxies at $1 \lesssim z \lesssim 3$. 

\end{itemize}



\acknowledgments

We thank an anonymous referee for very helpful comments.
This study is based on data collected at Subaru Telescope, which is operated by
the National Astronomical Observatory of Japan. 
This work is based in part on observations made with the Spitzer Space 
Telescope, which is operated by the Jet Propulsion Laboratory, 
California Institute of Technology under a contract with NASA.
Some of the data presented in this paper were obtained from the Multimission 
Archive at the Space Telescope Science Institute (MAST). 
STScI is operated by the Association of Universities for Research in 
Astronomy, Inc., under NASA contract NAS5-26555. 
Support for MAST for non-HST data is provided by the NASA Office of 
Space Science via grant NAG5-7584 and by other grants and contracts.
Data reduction/analysis was carried out on ``sb'' computer system
operated by the Astronomical Data Analysis Center (ADAC) and Subaru
Telescope of the National Astronomical Observatory of Japan.
IRAF is distributed by the National Optical Astronomy Observatories,
which are operated by the Association of Universities for Research
in Astronomy, Inc., under cooperative agreement with the National
Science Foundation.

\clearpage

\end{document}